\documentclass[12pt, onecolumn,draftcls]{IEEEtran} 

\usepackage{amsmath}
\usepackage{amsfonts}
\usepackage{multirow}
\usepackage{nicefrac}
\usepackage{cite}
\usepackage{epsfig}
\usepackage{color}
\usepackage{graphicx,psfrag}

\newtheorem{lemma}{Lemma}[section]
\newtheorem{theorem}{Theorem}[section]
\newtheorem{corollary}{Corollary}[section]
\newtheorem{proposition}{Proposition}[section]
\newtheorem{remark}{Remark}[section]
\newtheorem{example}{Example}[section]

\title{Complete Characterization of the Equivalent MIMO Channel for Quasi-Orthogonal Space-Time Codes}

\author{Aydin Sezgin\\
Fraunhofer-Institute for Telecomm., HHI \\
 Broadband Mobile Comm. Networks \\
  Einsteinufer 37, 10587 Berlin, Germany\\
 Phone/Fax: +49 30 31002 868/863 \\
aydin.sezgin@hhi.fhg.de \\ \thanks{This paper was presented in part at the ITW 2004, San Antonio, TX , USA, October
2004. This work was supported in part by the German ministry of education and research (BMBF) under grant
01BU150.} \and Tobias J. Oechtering \\
                                    Faculty of EECS, Technical University of Berlin\\
                                    Heinrich Hertz Chair for Mobile Comm. Technology\\
                                    Einsteinufer 25, 10587 Berlin, Germany\\
                                    Phone/Fax: +49 30 31428 165/320 \\
                                    tobias.oechtering@mk.tu-berlin.de}

\begin{document}
\maketitle

\begin{abstract}
Recently, a quasi-orthogonal space-time block code (QSTBC) capable of achieving a significant fraction of the outage
mutual information of a multiple-input-multiple output (MIMO) wireless communication system for the case of four
transmit and one receive antennas was proposed. We generalize these results to $n_T=2^n$ transmit and an arbitrary
number of receive antennas $n_R$. Furthermore, we completely characterize the structure of the equivalent channel for
the general case and show that for all $n_T=2^n$ and $n_R$ the eigenvectors of the equivalent channel are fixed and
independent from the channel realization. Furthermore, the eigenvalues of the equivalent channel are independent
identically distributed random variables each following a noncentral chi-square distribution with $4n_R$ degrees of
freedom. Based on these important insights into the structure of the QSTBC, we derive an analytical lower bound for the
fraction of outage probability achieved with QSTBC and show that this bound is tight for low signal-to-noise-ratios
(SNR) values and also for increasing number of receive antennas. We also present an upper bound, which is tight for
high SNR values and derive analytical expressions for the case of four transmit antennas. Finally, by utilizing the
special structure of the QSTBC we propose a new transmit strategy, which decouples the signals transmitted from
different antennas in order to detect the symbols separately with a linear ML-detector rather than joint detection, an
up to now only known advantage of orthogonal space-time block codes (OSTBC).
\end{abstract}

\section{Introduction}
In recent years, the goal of providing high speed wireless data services has generated a great amount of interest among
the research community. Recent information theoretic results have demonstrated that the ability of a system to support
a high link quality and higher data rates in the presence of Rayleigh fading improves significantly with the use of
multiple transmit and receive antennas~\cite{Telatar99,FoschiniGans98}. A very important aspect is always the
availability of analytical expressions to describe the stochastic nature of the channel under consideration as given
in~\cite{Telatar99,Edelman} for the MIMO channel. This offers an opportunity to obtain, e.g., closed-form analytical
formulas for the ergodic capacity or the outage mutual information of such MIMO channels. E.g., in~\cite{Giannakis},
the probability density function (pdf) of the random mutual information for independent identically distributed
(i.i.d.) MIMO channels was derived in the form of the inverse Laplace transform and a Gaussian approximation of the pdf
was presented. In~\cite{H.Shin2003}, the impact of MIMO channel rank deficiency and spatial fading correlation on the
mutual information was analyzed. Furthermore, for correlated channels, the optimal transmit strategy and the impact of
correlation on the outage probability were derived in~\cite{Jorswieck}.

There has been considerable work on a variety of new codes and modulation signals, called space-time codes, in order to
approach the huge capacity of such MIMO channels. The performance criteria of space-time codes were derived
in~\cite{GueyFitzBellKuo99,TarokhSeCa98}. One scheme of particular interest is the Alamouti scheme~\cite{Alamouti} for
two transmit antennas. Later on,~\cite{TarokhJafarkCalder99} proposed more general schemes referred to as orthogonal
space-time block codes (OSTBC) with the same properties as the Alamouti scheme like, e.g., a remarkably simple
maximum-likelihood decoding algorithm. Interestingly, the combination of OSTBC with a MIMO antenna system can be
represented equivalently as a single-input-single-output (SISO) system, where the channel gain is equal to the
Frobenius norm of the actual MIMO channel. The performance of orthogonal space-time block
codes~\cite{TarokhJafarkCa99,TirkkHottin,X.B.Liang2003} with respect to mutual information was analyzed (among others)
for the uncorrelated Rayleigh fading case in~\cite{Sandhu,Bauch} and for the more general case with different
correlation scenarios and line of sight (LOS) components in~\cite{Paulraj}. OSTBC exploit multiple antennas at both the
transmitter and receiver in order to obtain transmit and receive diversity and therefore increase the reliability of
the system. With the knowledge of the stochastic nature of the resulting equivalent channel due to the employment of
OSTBC in a MIMO system,
 the loss in mutual information of OSTBC in subject to transmission rate, number of receive antennas
and channel rank was quantified in~\cite{Sandhu}, whereas in~\cite{Bauch} a comparison of OSTBC with a system applying
beam-forming was presented.

Unfortunately, the Alamouti space-time code for two transmit and one receive antennas is the only OSTBC, which, to the
best of our knowledge, achieves the maximum possible mutual information of a MIMO system~\cite{Telatar99}, since we can
not construct an OSTBC with transmission rate equal one for more than two transmit
antennas~\cite{TarokhJafarkCalder99,LiangXia}. Therefore, \cite{Jafarkhani01,PapadiasFoschi01,TirkkonenHottinenNON}
designed a quasi-orthogonal space-time block code (QSTBC) with transmission rate one for four and eight transmit
antennas. By properly choosing the signal constellations as done
in~\cite{SharmaPapadias02,SharmaPapadias03,WeifengSuXia,WeifengSuXiaInfoTheory,SezginJorswieckBoche,TirkkonenHottinenImproved,SharmaPapadiasEURASIP},
it is possible to improve the BER performance with ML-detection for the codes given
in~\cite{Jafarkhani01,PapadiasFoschi01,TirkkonenHottinenNON}. The BER performance of QSTBC with suboptimal detectors
has been analyzed in~\cite{MecklenbraeuckerGLOBE,MecklenbraeuckerWPMC}.

The performance of QSTBC with respect to outage mutual information (OMI) for the special case of one receive antenna
and four or eight transmit antennas was analyzed via simulations in~\cite{PapadiasFoschi01}
and~\cite{C.F.Mecklenbrauker2004} and it was shown, that the QSTBC are capable to achieve a significant portion of the
MIMO-OMI. Furthermore, it was shown in~\cite{PapadiasFoschi01}, that QSTBC in conjunction with optimal (nonlinear) and
suboptimal (linear) detectors provide a tradeoff between performance and complexity. The key achievements of this paper
are as follows
\begin{itemize}
    \item We generalize the results in~\cite{PapadiasFoschi01} to $2^n$ transmit and
an arbitrary number of receive antennas.
\item We show, that due to the employment of QSTBC the eigenvalues of the resulting equivalent channel are pairwise independent and
identical (i.i.d) noncentral chi-square distributed with $4n_R$ degrees of freedom ($\chi_{4n_R}^2(\delta_{nc})$) with
noncentrality parameter $\delta_{nc}$. Furthermore, we show that the eigenvectors of the equivalent channel are
independent of each channel realization, i.e. they are constant.
\end{itemize}
In other words, we first show that the combination of QSTBC with a MIMO system results also in a equivalent channel
similar to OSTBC and then fully characterize the stochastic nature of this equivalent channel. Based on these important
insights, we are able to provide the following results
\begin{itemize}
    \item an analytical lower bound for the outage
probability achieved with QSTBC, which is tight for low signal-to-noise-ratios (SNR) values and also for increasing
number of receive antennas;
    \item an upper bound on the outage probability, which is tight for high SNR values. For the case of four transmit
    and an arbitrary number of receive antennas we derive analytical expressions for this bound.
    \item finally, we exploit the special
structure of the QSTBC and apply a new transmit strategy, which decouples the signals transmitted from different
antennas in order to detect the symbols separately as in the case of OSTBC. The performance of this linear
detector is equivalent to the non-linear maximum-likelihood (ML)-detector in~\cite{PapadiasFoschi01}.
\end{itemize}

 The remainder of this paper is organized as follows. In Section~\ref{seq:system_model},
we introduce the system model and establish the notation. The design of QSTBC for $2^n$ transmit antennas is shown in
section~\ref{sec:Diversity_analy}. The complete characterization of the equivalent channel model and other key
achievements of this paper are described in section~\ref{sec:Preprocessing}. As an possible application of the results
in this section, we present the analysis of the outage probability achieved with QSTBC in section~\ref{sec:PerfAnaly},
followed by some simulations and concluding remarks in section~\ref{seq:simulations} and~\ref{seq:Conclusion}.

\section{System model}\label{seq:system_model}
 We consider a system with $n_T=2^n$ transmit and $n_R$ receive antennas. Our system model is defined
by
\begin{equation}\label{eq:System}
    \mathbf{Y} = \mathbf{G}_{n_T}\mathbf{H}+  \mathbf{N}\;,
\end{equation}
where $\mathbf{G}_{n_T}$ denotes the ($T \times n_T $) transmit matrix,
$\mathbf{Y}=[\mathbf{y}_1,\dots,\mathbf{y}_{n_R}]$ the ($T \times n_R$) receive matrix,
$\mathbf{H}=[\mathbf{h}_1,\dots,\mathbf{h}_{n_R}]$ the ($n_T \times n_R $) channel matrix, and
$\mathbf{N}=[\mathbf{n}_1,\dots,\mathbf{n}_{n_R}]$  the complex ($T \times n_R$) white Gaussian noise (AWGN) matrix,
respectively. An entry $\{n_{ti}\}$ of $\mathbf{N}$ ($1\leq i \leq n_R$) denotes the complex noise  at the $i$th
receiver for a given time instant $t\,(1\leq t \leq T)$. The real and imaginary parts of $n_{ti}$ are independent and
$\mathcal{N}$(0,$n_T/(2\mathrm{SNR})$) distributed. An entry of the channel matrix is represented by $\{h_{ji}\}\in
\mathbf{h}_i$ and describes the complex gain of the channel between the $j$th transmit ($1\leq j \leq n_T$) and the
$i$th receive ($1\leq i \leq n_R$) antenna, where the real and imaginary parts of the channel gains are independent and
normal distributed random variables and $\mathbf{h}_i$ is $\mathcal{CN}(\mathbf{m}_i,\mathbf{I})$ distributed, where
$\mathbf{m}_i$ is the channel mean or Ricean component. The channel matrix is assumed to be constant for a block of $T$
symbols and changes independently from block to block. The average power of the symbols transmitted from each antenna
is normalized to one, so that the average power of the received signal at each receive antenna is $n_T$ and the
signal-to-noise ratio (SNR) is $\rho$. It is further assumed that the transmitter has no CSI and the receiver has
perfect CSI.

\section{Code construction}\label{sec:Diversity_analy}
A space-time block code is defined by its transmit matrix $\mathbf{G}_{n_T}$, which is a function of the vector
$\mathbf{x}=[x_1,\dots,x_p]^T$. The rate $R$ of a space-time block code is defined as $R=p/T$. In this paper, we focus
on rate one QSTBC with length $n_T=T$, therefore $p=n_T$. Now, let us split the vector $\mathbf{x}$ into two vectors,
$\mathbf{x}_{\mathrm{odd}}$ and $\mathbf{x}_{\mathrm{even}}$, for reasons that will be clear later on. The elements of
$\mathbf{x}$ with odd index $j$ are collected in $\mathbf{x}_{\mathrm{odd}}$ and with even index in
$\mathbf{x}_{\mathrm{even}}$, respectively. Both parts of $\mathbf{x}$ are given as
\begin{equation}\label{eq:x_oddANDx_evenDEF}
 \mathbf{x}_{\mathrm{odd}} = \mathbf{\Gamma}\left[\begin{array}{c}
   s_1 \\
   \vdots \\
   s_{\nicefrac{n_T}{2}} \\
 \end{array}\right]= \mathbf{\Gamma}\mathbf{s}^-,
    \mathbf{x}_{\mathrm{even}}=  \mathbf{\Gamma}\left[\begin{array}{c}
   s_{\nicefrac{n_T}{2}+1} \\
   \vdots \\
   s_{n_T} \\
 \end{array}\right]= \mathbf{\Gamma}\mathbf{s}^+,
\end{equation}
with $s_1,\dots,s_{n_T} \in \mathcal{C}$, where $\mathcal{C} \subseteq \mathbb{C}$ denotes a complex modulation
signal set with unit average power, e.g. $M$-PSK. Furthermore, $\mathbf{\Gamma} \in
\mathbb{C}^{\nicefrac{n_T}{2}\times \nicefrac{n_T}{2}}$ is a unitary matrix. More details on $\mathbf{\Gamma}$ and
its effect on the detection scheme will be discussed in section~\ref{sec:trans_strat}.

Starting with the well known Alamouti scheme~\cite{Alamouti} for $n_T=2$ transmit antennas as a
\begin{equation}\nonumber
 \mathbf{G}_{2}(x_1,x_2) = \left[
\begin{array}{*{2}{r}}
          x_1 & x_2  \\
        x_2^* & -x_1^*  \\
        \end{array}
\right]\;,
\end{equation}
the generalization of the transmit matrix for the QSTBC with $n_T=2^n$ ($n_T\geq4$) is done in the following recursive
way
\begin{equation}
\mathbf{G}_{n_T}\left(\{x_j\}_{j=1}^{n_T}\right) = \left[
\begin{array}{*{2}{c}}
          \mathbf{G}_{\frac{n_T}{2}}\left(\{x_j\}_{j=1}^{\frac{n_T}{2}}\right) & \mathbf{G}_{\frac{n_T}{2}}\left(\{x_j\}_{j=\frac{n_T}{2}+1}^{n_T}\right)  \\
        \mathbf{G}_{\frac{n_T}{2}}\left(\{x_j\}_{j=\frac{n_T}{2}+1}^{n_T}\right)\mathbf{\Theta}_{n_T} & -\mathbf{G}_{\frac{n_T}{2}}\left(\{x_j\}_{j=1}^{\frac{n_T}{2}}\right)\mathbf{\Theta}_{n_T}  \\
        \end{array}
\right]\nonumber\;,
\end{equation}
where $\{x_j\}_{j=1}^{n_T}=x_1,\dots,x_{n_T}$ and the diagonal $\nicefrac{n_T}{2} \times \nicefrac{n_T}{2}$ matrix
$\mathbf{\Theta}_{n_T}$ is given by
$\mathbf{\Theta}_{n_T}=\mathrm{diag}\left(\{(-1)^{j-1}\}_{j=1}^{\frac{n_T}{2}}\right)$.

\begin{example}
For the case of $n_T=4$ transmit antennas we have
\begin{equation}\nonumber
\mathbf{G}_{4}(\{x_j\}_{j=1}^{4}) = \left[
\begin{array}{*{4}{r}}
          x_1 & x_2 & x_3 & x_4 \\
        x_2^* & -x_1^* & x_4^* & -x_3^* \\
        x_3 & -x_4 & -x_1 & x_2 \\
       x_4^* & x_3^* & -x_2^* & -x_1^* \\
        \end{array}
\right]\;.
\end{equation}

\end{example}
In this work, we use the Alamouti scheme as the basis in order to construct the rate one QSTBC. However, it is also
possible to construct QSTBC with rates lower than one based on other OSTBC~\cite{TarokhJafarkCalder99,TirkkHottin}. In
the following section, we perform channel-matched filtering as the first stage of preprocessing at the receiver in
order to obtain the equivalent channel model, followed by the decoupling of the system model in two parts. Afterwards,
we analyze the eigenvalues and the eigenvectors of the resulting equivalent channel, leading to important insights of
the properties of QSTBC. Noise pre-whitening as the second stage of preprocessing at the receiver is considered in
section~\ref{sec:Noise pre-whitening}.

\section{Signal Processing}\label{sec:Preprocessing}
First of all, we briefly review the usual MIMO fading channel without any coordinated coding and the impact of OSTBC on
the MIMO channel in order to provide a better insight into the properties of QSTBC.
\subsection{MIMO channel without any coordinated coding}
In this case, after channel matched filtering to (\ref{eq:System}), we have
\begin{equation}\label{eq:SVD_UsualMIMO}
\mathbf{H}\mathbf{H}^H=\mathbf{V}\mathbf{D}\mathbf{D}\mathbf{V}^H\;,
\end{equation}
where $\mathbf{H}=\mathbf{VDU}^H$ is the singular value decomposition (SVD) of $\mathbf{H}$, where the unitary matrices
$\mathbf{U}$,$\mathbf{V}$ contain the eigenvectors of $\mathbf{H}$. The joint density function of the eigenvalues
$\mu_1,\dots,\mu_m$ of $\mathbf{H}\mathbf{H}^H$ in $\mathbf{D}\mathbf{D}$ in the Rayleigh fading case
$(\mathbf{m}_i=\mathbf{0})$ is given as~\cite{Telatar99,Edelman}
\begin{equation}
p_\mathbf{\mu}(\mu_1,\dots,\mu_m)=\frac{1}{m!K_{m,n}}\;e^{\sum_i\mu_i}\prod_i\mu_i^{n-m}\prod_{i<j}(\mu_i-\mu_j)^2\;,
\end{equation}
where $K_{m,n}$ is a normalizing factor, $n=\max\{n_T,n_R\}$ and $m=min\{n_T,n_R\}$. It is obvious, that the
eigenvalues are not independent of each other and it is well known that the matrix of eigenvectors $\mathbf{V}$ depend
on the actual channel realization.
\subsection{Equivalent channel for OSTBC} In case of OSTBC, the following holds for the
transmit matrix
\begin{equation}\nonumber
    \mathbf{G}_{n_T}^H\mathbf{G}_{n_T}=\sum\limits_{j=1}^{p}|x_j|^2\mathbf{I}_{n_T}\;.
\end{equation}
Starting with (\ref{eq:System}), after some manipulations and channel matched filtering one arrives at
\begin{equation}\nonumber
    \mathbf{y}''=\mathbf{H}_{n_T}''\mathbf{x}+ \mathbf{n}''\;,
\end{equation}
where
\begin{equation}
    \mathbf{H}_{n_T}''=\left[%
\begin{array}{ccc}
  \sum\limits_{i=1}^{n_R}\sum\limits_{j=1}^{n_T}|h_{ji}|^2 &  & $\large{0}$ \\
   & \ddots &  \\
  $\large{0}$ &  & \sum\limits_{i=1}^{n_R}\sum\limits_{j=1}^{n_T}|h_{ji}|^2  \\
\end{array}%
\right].
\end{equation}
Since there is no interaction between the elements of $\mathbf{x}$, the equation above can be decomposed into $p$
parts. The resulting equivalent channel for each element of $\mathbf{x}$ of the OSTBC is then a
single-input-single-output (SISO) channel given as
\begin{equation}
    \widetilde{\mathbf{H}}_{\frac{n_T}{p}}=\sum\limits_{i=1}^{n_R}\sum\limits_{j=1}^{n_T}|h_{ji}|^2
\end{equation}
 which is equal to the Frobenius norm of the actual MIMO
channel matrix $\mathbf{H}$.

In case of the rate one QSTBC discussed in this paper, the actual MIMO channel is also transformed into a equivalent
channel given as $\widetilde{\mathbf{H}}_{\frac{n_T}{2}}$. Differently from the OSTBC the equivalent channel of QSTBC
is still a MIMO channel, however with very interesting properties like constant eigenvectors and i.i.d. eigenvalues
following a noncentral $\chi_{4n_R}^2(\delta_{nc})$-distribution as derived in the following.

\subsection{Channel-Matched Filtering}
After rearranging and complex-conjugating some rows of $\mathbf{Y}$ the system equation in (\ref{eq:System}) can be
rewritten as
\begin{equation}\label{eq:system_H_vorne}
    \mathbf{y}'=\mathbf{H}_{n_T}'\mathbf{x}+ \mathbf{n}'\;,
\end{equation}
where $\mathbf{H}_{n_T}'=[(\mathbf{H}_{n_T,1}')^T,\dots,(\mathbf{H}_{n_T,i}')^T,\dots,(\mathbf{H}_{n_T,n_R}')^T]^T$ and
$\mathbf{H}_{n_T,i}'$ is given as
\begin{equation}\label{eq:neuer_Kanal}
\mathbf{H}_{n_T,i}'= \mathbf{H}_{n_T,i}'\left(\{h_{ji}\}_{j=1}^{n_T}\right) =\left[
\begin{array}{*{2}{c}}
           \mathbf{H}_{\frac{n_T}{2}}\left(\{h_{ji}\}_{j=1}^{\frac{n_T}{2}}\right)&   \mathbf{H}_{\frac{n_T}{2}}\left(\{h_{ji}\}_{j=\frac{n_T}{2}+1}^{n_T}\right) \\
        -\mathbf{\Theta}_{n_T}\mathbf{H}_{\frac{n_T}{2}}\left(\{h_{ji}\}_{j=\frac{n_T}{2}+1}^{n_T}\right)\mathbf{\Theta}_{n_T} & \mathbf{\Theta}_{n_T}\mathbf{H}_{\frac{n_T}{2}}\left(\{h_{ji}\}_{j=1}^{\frac{n_T}{2}}\right)\mathbf{\Theta}_{n_T}  \\
        \end{array}
\right].
\end{equation}
Thus it appears that the $\mathbf{H}_{n_T,i}'\left(\{h_{ji}\}_{j=1}^{n_T}\right)$, with
$\left(\{h_{ji}\}_{j=1}^{n_T}\right)=h_{1i},\dots,h_{n_Ti}$, are obtained recursively, where the recursion starts with
$n_T=2$,
\begin{equation}\nonumber
 \mathbf{H}_{2,i}=\mathbf{H}_{2,i}(h_{1i},h_{2i}) = \left[
\begin{array}{*{2}{r}}
          h_{1i} & h_{2i}  \\
        -h_{2i}^* & h_{1i}^*  \\
        \end{array}
\right]\;.
\end{equation}
In order to perform channel-matched filtering we multiply $(\mathbf{H}_{n_T}')^H$ from left to
(\ref{eq:system_H_vorne}) to get
\begin{equation}\label{eq:system_zerlegt}
    \mathbf{y}''=\mathbf{H}_{n_T}''\mathbf{x}+ \mathbf{n}''\;,
\end{equation}
where the noise vector $\mathbf{n}''=(\mathbf{H}_{n_T}')^H\mathbf{n}'$ is spatially colored and $\mathbf{H}_{n_T}''$ is
given as
\begin{equation}\label{eq:Channel_VorZerleg}
\mathbf{H}_{n_T}''= \left[%
\begin{array}{cc}
  \mathbf{K}^H \mathbf{K}+\mathbf{L}^H \mathbf{L}                                               & \mathbf{\Theta}_{n_T}\mathbf{K}^H \mathbf{L}\mathbf{\Theta}_{n_T}-\mathbf{L}^H \mathbf{K} \\
   -(\mathbf{\Theta}_{n_T}\mathbf{K}^H \mathbf{L}\mathbf{\Theta}_{n_T}-\mathbf{L}^H \mathbf{K}) & \mathbf{K}^H \mathbf{K}+\mathbf{L}^H \mathbf{L} \\
\end{array}%
\right],
\end{equation}
where $\mathbf{K}=\mathbf{H}_{\frac{n_T}{2}}\left(\{h_{ji}\}_{j=1}^{\frac{n_T}{2}}\right)$ and
$\mathbf{L}=\mathbf{H}_{\frac{n_T}{2}}\left(\{h_{ji}\}_{j=\frac{n_T}{2}+1}^{n_T}\right)$.
\subsection{Decoupling of the system model}
An important property of the QSTBC the system in (\ref{eq:system_zerlegt}) can be decoupled into two parts due to the
special structure of $\mathbf{H}_{n_T}''$ as described in the following. The decoupling comes from the fact that for
QSTBC, it holds that~\cite{SharmaPapadias02}
\begin{equation}\label{eq:ZerlegProbQSTBC}
\mathbf{G}_{n_T}^H(\tilde{\mathbf{x}}_{\mathrm{odd}}) \cdot  \mathbf{G}_{n_T}(\tilde{\mathbf{x}}_{\mathrm{even}})  +
\mathbf{G}_{n_T}^H(\tilde{\mathbf{x}}_{\mathrm{even}}) \cdot \mathbf{G}_{n_T}(\tilde{\mathbf{x}}_{\mathrm{odd}})
=\mathbf{0} \quad\forall\, { \mathbf{x}} \;,
\end{equation}
where $\tilde{\mathbf{x}}_{\mathrm{odd}}=\mathbf{x}_{\mathrm{odd}}\otimes [%
\begin{array}{cc}
  1 & 0 \\
\end{array}%
]^T=[x_1,0,x_3,0,\dots,x_{n_T-1},0]^T$ and $\tilde{\mathbf{x}}_{\mathrm{even}}=\mathbf{x}_{\mathrm{even}}\otimes [%
\begin{array}{cc}
  0 & 1 \\
\end{array}%
]^T$. The property in~(\ref{eq:ZerlegProbQSTBC}) is very crucial, because this enables a simple maximum-likelihood
decoding algorithm. Assuming perfect channel estimation is available, the receiver computes the following decision
metric over all possible transmit matrices and decides in favor of the transmit matrix that minimizes the following
decision metric based on~(\ref{eq:System}):
\begin{eqnarray}\label{eq:maximum_liklihood}
  ||\mathbf{Y}- \mathbf{G}_{n_T}(\mathbf{x})\cdot \mathbf{H} ||_F^2
  & = & \mathrm{tr} \{  (\mathbf{Y}- \mathbf{G}_{n_T}(\mathbf{x})\cdot \mathbf{H})^H(\mathbf{Y}-\mathbf{G}_{n_T}(\mathbf{x})\cdot \mathbf{H} ) \} \\
 &=  & \mathrm{tr} \{\mathbf{Y}^H\mathbf{Y}-\mathbf{Y}^H\mathbf{G}_{n_T}(\mathbf{x})\mathbf{H}  \nonumber \\
& & -
(\mathbf{Y}^H\mathbf{G}_{n_T}(\mathbf{x})\mathbf{H})^H+\mathbf{H}^H\mathbf{G}_{n_T}(\mathbf{x})^H\mathbf{G}_{n_T}(\mathbf{x})\mathbf{H}\}
\nonumber \;.
\end{eqnarray}
After some manipulations, we arrive at
\begin{equation}\nonumber
\begin{split}
\mathrm{tr} \{  \mathbf{Y}_{\mathrm{odd}}^H\mathbf{Y}_{\mathrm{odd}}
+\mathbf{Y}_{\mathrm{odd}}^H\mathbf{G}_{n_T}(\tilde{\mathbf{x}}_{\mathrm{odd}})\mathbf{H}
+\mathbf{H}^H\mathbf{G}_{n_T}^H(\tilde{\mathbf{x}}_{\mathrm{odd}})\mathbf{Y}_{\mathrm{odd}}
+\mathbf{H}^H\mathbf{G}_{n_T}^H(\tilde{\mathbf{x}}_{\mathrm{odd}})\mathbf{G}_{n_T}(\tilde{\mathbf{x}}_{\mathrm{odd}})\mathbf{H}+  \\
    \mathbf{Y}_{\mathrm{even}}^H\mathbf{Y}_{\mathrm{even}} +\mathbf{Y}_{\mathrm{even}}^H\mathbf{G}_{n_T}(\tilde{\mathbf{x}}_{\mathrm{even}})\mathbf{H} +\mathbf{H}^H\mathbf{G}_{n_T}^H(\tilde{\mathbf{x}}_{\mathrm{even}})\mathbf{Y}_{\mathrm{even}}
    +\mathbf{H}^H\mathbf{G}_{n_T}^H(\tilde{\mathbf{x}}_{\mathrm{even}})\mathbf{G}_{n_T}(\tilde{\mathbf{x}}_{\mathrm{even}})\mathbf{H}\} \;,
\end{split}
\end{equation}
where $\mathrm{tr}\{\cdot\}$ is the trace function. $\mathbf{Y}_{\mathrm{odd}}$ and $\mathbf{Y}_{\mathrm{even}}$ are
given as
\begin{equation}
  \mathbf{Y}_{\mathrm{odd}} = \mathbf{G}_{n_T}(\tilde{\mathbf{x}}_{\mathrm{odd}})\mathbf{H}+\mathbf{N}_{\mathrm{odd}}\; \text{ and }\;
  \mathbf{Y}_{\mathrm{even}} = \mathbf{G}_{n_T}(\tilde{\mathbf{x}}_{\mathrm{even}})\mathbf{H}+\mathbf{N}_{\mathrm{even}} \nonumber
  \;,
\end{equation}
respectively. The above decision metric can be decomposed into two parts, one of which
\begin{equation}\nonumber
\mathrm{tr} \{ \mathbf{Y}_{\mathrm{odd}}^H\mathbf{Y}_{\mathrm{odd}}
+\mathbf{Y}_{\mathrm{odd}}^H\mathbf{G}_{n_T}(\tilde{\mathbf{x}}_{\mathrm{odd}})\mathbf{H}
+\mathbf{H}^H\mathbf{G}_{n_T}^H(\tilde{\mathbf{x}}_{\mathrm{odd}})\mathbf{Y}_{\mathrm{odd}}
+\mathbf{H}^H\mathbf{G}_{n_T}^H(\tilde{\mathbf{x}}_{\mathrm{odd}})\mathbf{G}_{n_T}(\tilde{\mathbf{x}}_{\mathrm{odd}})\mathbf{H}\}
\end{equation}
is only a function of $\mathbf{G}_{n_T}(\tilde{\mathbf{x}}_{\mathrm{odd}})$, and the other one
\begin{equation}\nonumber
       \mathrm{tr} \{\mathbf{Y}_{\mathrm{even}}^H\mathbf{Y}_{\mathrm{even}} +\mathbf{Y}_{\mathrm{even}}^H\mathbf{G}_{n_T}(\tilde{\mathbf{x}}_{\mathrm{even}})\mathbf{H} +\mathbf{H}^H\mathbf{G}_{n_T}^H(\tilde{\mathbf{x}}_{\mathrm{even}})\mathbf{Y}_{\mathrm{even}}
    +\mathbf{H}^H\mathbf{G}_{n_T}^H(\tilde{\mathbf{x}}_{\mathrm{even}})\mathbf{G}_{n_T}(\tilde{\mathbf{x}}_{\mathrm{even}})\mathbf{H}\}\;,
\end{equation}
is only a function of $\mathbf{G}_{n_T}(\tilde{\mathbf{x}}_{\mathrm{even}})$. Thus the minimization of
(\ref{eq:maximum_liklihood}) is equivalent to minimizing these two parts separately. Note that, due to the processing
at the receiver, the property in~(\ref{eq:ZerlegProbQSTBC}) is projected on the channel matrix $\mathbf{H}_{n_T}''$
in~(\ref{eq:system_zerlegt}). The decoupled parts depend either on $\mathbf{x}_{\mathrm{odd}}$ or
$\mathbf{x}_{\mathrm{even}}$ given in (\ref{eq:x_oddANDx_evenDEF}).

Thus, it is now possible to write an decomposed system model for each part based on~(\ref{eq:system_zerlegt}). The
decomposed system model for the part with $\mathbf{x}_{\mathrm{odd}}$ (and similarly for $\mathbf{x}_{\mathrm{even}}$)
can be rewritten as
\begin{equation}\label{eq:system_model_s1s3_colored_noise}
    \mathbf{y}_{\mathrm{odd}}=\widetilde{\mathbf{H}}_{\frac{n_T}{2}} \mathbf{x}_{\mathrm{odd}} + \widetilde{\mathbf{n}}\;.
\end{equation}
For illustration, we present two examples for the case of $n_T=4$ and $n_T=8$ transmit antennas.

\begin{example}($n_T=4$ transmit antennas)
In this case, $\mathbf{H}_{4,i}'$ in (\ref{eq:neuer_Kanal}) is given as
\begin{equation}
\mathbf{H}_{4,i}'=\left[%
\begin{array}{rrrr}
   h_{1i}   &  h_{2i}   &   h_{3i}   &  h_{4i}   \\
  -h_{2i}^* &  h_{1i}^* &  -h_{4i}^* &  h_{3i}^* \\
  -h_{3i}   &  h_{4i}   &   h_{1i}   & -h_{2i}   \\
  -h_{4i}^* & -h_{3i}^* &   h_{2i}^* &  h_{1i}^* \\
\end{array}%
\right] \nonumber \;.
\end{equation}
and $\mathbf{H}_{4}''$ appears in (\ref{eq:system_zerlegt}) as
\begin{equation}\label{eq:system_zerlegbar_Bsp4MT}
    \mathbf{y}''=\left[%
\begin{array}{cccc}
    \alpha_1  & 0 &   i\alpha_2 & 0 \\
  0 & \alpha_1& 0 & -i\alpha_2 \\
  -i\alpha_2   & 0&   \alpha_1 & 0 \\
  0 & i\alpha_2&   0 & \alpha_1 \\
\end{array}%
\right]\left[\begin{array}{c}
      x_1 \\
      x_2 \\
      x_3 \\
      x_4 \\
    \end{array}\right]+ \mathbf{n}''\;,
\end{equation}
where $\alpha_1$ and $\alpha_2$ are given as
\begin{equation}\label{eq:alpha1u2forNt4}
    \alpha_1 = \sum_{i=1}^{n_R}\sum_{j=1}^{4} |h_{ji}|^2\; \text{ and } \alpha_2 =  \sum_{i=1}^{n_R}2\mathrm{Im}(h_{1i}^*h_{3i}+h_{4i}^*h_{2i}),
\end{equation}
respectively. From (\ref{eq:system_zerlegbar_Bsp4MT}), it is now directly obvious, that the system equation can be
decoupled into two parts, which then can be considered separately. For the case considered in this example, the
decomposed system model for  $\mathbf{x}_{\mathrm{odd}}$ (and similarly for $\mathbf{x}_{\mathrm{even}}$, cf.
(\ref{eq:system_model_s1s3_colored_noise})) can be written as
\begin{equation}\nonumber
    \mathbf{y}_{\mathrm{odd}}=\widetilde{\mathbf{H}}_{2} \left[\begin{array}{c}
      x_1 \\
      x_3 \\
    \end{array}\right] + \widetilde{\mathbf{n}}\;,
\end{equation}
with a non-orthogonal
\begin{equation}\label{eq:Kanal_Xodd}
\widetilde{\mathbf{H}}_{2}=\left[%
\begin{array}{cc}
    \alpha_1   &   i\alpha_2 \\
  -i\alpha_2 & \alpha_1 \\
\end{array}%
\right] \;.
\end{equation}
\end{example}

\begin{example}($n_T=8$ transmit antennas)
The same procedure applied here results in a $\widetilde{\mathbf{H}}_{4}$ given as
\begin{equation}\nonumber
\widetilde{\mathbf{H}}_{\frac{n_T}{2}}=\widetilde{\mathbf{H}}_{4}=\left[\begin{array}{cccc}
   \alpha_1 &  i\alpha_2  &  i\alpha_3   & \alpha_4  \\
  -i\alpha_2  &  \alpha_1 & -\alpha_4  & i\alpha_3   \\
  -i\alpha_3   & -\alpha_4  &  \alpha_1 & i\alpha_2  \\
   \alpha_4  & -i\alpha_3   & -i\alpha_2  & \alpha_1 \\
\end{array}\right]\;,
\end{equation}
where
\begin{eqnarray}\label{eq:alpha1bis4Nt8}
\begin{split}
    \alpha_1 &= \sum_{i=1}^{n_R}\sum_{j=1}^{8} |h_{ji}|^2,\; \alpha_2 =
\sum_{i=1}^{n_R}2\mathrm{Im}(h_{1i}^*h_{3i}+h_{4i}^*h_{2i}+h_{5i}^*h_{7i}+h_{8i}^*h_{6i}),\\
 \alpha_3 &= \sum_{i=1}^{n_R}2\mathrm{Im}(h_{1i}^*h_{5i}+h_{6i}^*h_{2i}+h_{3i}^*h_{7i}+h_{8i}^*h_{4i}),\text{ and } \\
 \alpha_4 &= \sum_{i=1}^{n_R}2\mathrm{Re}(h_{1i}^*h_{7i}+h_{8i}^*h_{2i}-h_{3i}^*h_{5i}-h_{6i}^*h_{4i}).
\end{split}
\end{eqnarray}
\end{example}
The general case of arbitrary $n_T=2^n$ and very important insights about the eigenvalue decomposition, the eigenvalues
themselves and the eigenvectors of the equivalent channel $\widetilde{\mathbf{H}}_{\frac{n_T}{2}}$, which are crucial
and necessary for further analysis, e.g., the derivations of the lower and upper bound, are provided in the following
section.

\subsection{Eigenvalues and Eigenvectors of the equivalent channel model}\label{sec:EigValAndEigVecEquiChannel}
In order to completely characterize the equivalent channel with respect to the eigenvalues and eigenvectors, we first
look at the properties of matrices with certain structures and then show that the equivalent channel matrices fulfill
this special structure. Let the matrix $\mathbf{M}_N$, where $N=\frac{n_T}{2}=2^{n-1}$, has the following recursive
definition
\begin{equation}\label{eq:StructureA_N}
    \mathbf{M}_{N}(\alpha_1,\dots,\alpha_N)=\left[%
\begin{array}{rr}
  \mathbf{M}_{\frac{N}{2}}(\alpha_1,\dots,\alpha_{\frac{N}{2}}) & \mathbf{N}_{\frac{N}{2}}(\alpha_{\frac{N}{2}+1},\dots,\alpha_N) \\
  -\mathbf{N}_{\frac{N}{2}}(\alpha_{\frac{N}{2}+1},\dots,\alpha_N) & \mathbf{M}_{\frac{N}{2}}(\alpha_1,\dots,\alpha_{\frac{N}{2}}) \\
\end{array}%
\right]\;.
\end{equation}
Similarly
\begin{equation}\label{eq:StructureB_N}
    \mathbf{N}_{N}(\alpha_{N+1},\dots,\alpha_{2N})=\left[%
\begin{array}{rr}
  \mathbf{N}_{\frac{N}{2}}(\alpha_{N+1},\dots,\alpha_{\frac{3N}{2}}) & \mathbf{M}_{\frac{N}{2}}(\alpha_{\frac{3N}{2}+1},\dots,\alpha_{2N}) \\
  -\mathbf{M}_{\frac{N}{2}}(\alpha_{\frac{3N}{2}+1},\dots,\alpha_{2N}) & \mathbf{N}_{\frac{N}{2}}(\alpha_{N+1},\dots,\alpha_{\frac{3N}{2}}) \\
\end{array}%
\right]\;,
\end{equation}
where the recursion starts with
\begin{equation}\nonumber
    \mathbf{M}_2(\alpha_1,\alpha_2)=\left[%
\begin{array}{cc}
  \alpha_1 & i\alpha_2 \\
  -i\alpha_2 & \alpha_1 \\
\end{array}%
\right] \;\textrm{  and  }\;
    \mathbf{N}_2(\alpha_3,\alpha_4)=\left[%
\begin{array}{cc}
  i\alpha_3 & \alpha_4 \\
  -\alpha_4 & i\alpha_3 \\
\end{array}%
\right]\;.
\end{equation}
\begin{remark}
The matrices $\mathbf{M}_2$ and $\mathbf{N}_2$ have the following eigenvalue decompositions
\begin{equation}\label{eq:SVD_A_2B_2}
    \mathbf{M}_2=\mathbf{V}_2\mathbf{S}_2\mathbf{V}_2^H\;\textrm{ and }\;\mathbf{N}_2=\mathbf{V}_2\mathbf{T}_2\mathbf{V}_2^H
\end{equation}
where
\begin{equation}\nonumber
    \mathbf{V}_2=\frac{1}{\sqrt{2}}\left[%
\begin{array}{cc}
  1 & 1 \\
  -i & i \\
\end{array}%
\right]
\end{equation}
and
\begin{equation}\nonumber
 \mathbf{S}_2\left(\{\alpha_l\}_{l=1}^{2}\right)=\left[%
\begin{array}{cc}
  \mu_2^1 & 0 \\
  0 & \mu_2^2 \\
\end{array}%
\right]
=\phantom{i}\left[%
\begin{array}{cc}
  \alpha_1+\alpha_2 & 0 \\
  0 & \alpha_1-\alpha_2 \\
\end{array}%
\right]
\end{equation}
\begin{equation}\nonumber
\mathbf{T}_2\left(\{\alpha_l\}_{l=3}^{4}\right)=\left[%
\begin{array}{cc}
  \nu_2^1 & 0 \\
  0 & \nu_2^2 \\
\end{array}%
\right]=i\left[%
\begin{array}{cc}
  \alpha_3-\alpha_4 & 0 \\
  0 & \alpha_3+\alpha_4 \\
\end{array}%
\right]
\end{equation}
\end{remark}
Immediately the following question follows: Is their any structure how to derive the eigenvalues of the matrices of
higher $N$, i.e., if the eigenvalues of $\mathbf{M}_{\frac{N}{2}}$ are given, how can we compute the eigenvalues of
$\mathbf{M}_{N}$. (Note that, if the eigenvalues of $\mathbf{M}_{\frac{N}{2}}$ are given it is straightforward to
derive the eigenvalues of $\mathbf{N}_{\frac{N}{2}}$). In order to answer this question we are able to state the
following lemma, where the arguments of $\mathbf{M}_N$ and $\mathbf{N}_N$ are omitted.
\begin{lemma}\label{lem:FixedEigenv}
Let $\mathbf{M}_N,\mathbf{N}_N$ be as given in (\ref{eq:StructureA_N}),(\ref{eq:StructureB_N}), then
$\mathbf{M}_N,\mathbf{N}_N$ with $N=2^{n-1}$,$n> 2$ have the following eigenvalue decomposition:
\begin{equation}\label{eq:SVD_Statement}
    \mathbf{M}_N=\mathbf{V}_N\mathbf{S}_N\mathbf{V}_N^H\; \text{ and } \;
    \mathbf{N}_N=\mathbf{V}_N\mathbf{T}_N\mathbf{V}_N^H\;,
\end{equation}
where
\begin{equation}\label{eq:EigenvecVN}
    \mathbf{V}_N=
  \left(\mathbf{I}_2 \otimes\mathbf{V}_{\frac{N}{2}}\right)\mathbf{\Pi}_N\left(\mathbf{I}_{\frac{N}{2}} \otimes \mathbf{V}_2 \right)\;,
\end{equation}
\begin{equation}\label{eq:EigenvalSN}
\begin{aligned}
    \mathbf{S}_N =&\mathbf{S}_N\left(\{\alpha_l\}_{l=1}^{N}\right)=\mathbf{\Pi}_N \\
   &\cdot\left[%
\begin{array}{cc}
  \mathbf{S}_{\frac{N}{2}}\left(\{\alpha_l\}_{l=1}^{N/2}\right)-i\mathbf{T}_{\frac{N}{2}}\left(\{\alpha_l\}_{l=N/2+1}^{N}\right) & \mathbf{0} \\
  \mathbf{0} & \mathbf{S}_{\frac{N}{2}}\left(\{\alpha_l\}_{l=1}^{N/2}\right)+i\mathbf{T}_{\frac{N}{2}}\left(\{\alpha_l\}_{l=N/2+1}^{N}\right) \\
\end{array}%
\right]\mathbf{\Pi}_N^H\;,
\end{aligned}
\end{equation}
\begin{equation}\nonumber
\begin{aligned}
    &\mathbf{T}_N=\mathbf{T}_N\left(\{\alpha_l\}_{l=N+1}^{2N}\right)=\mathbf{\Pi}_N \\
    &\cdot\left[%
\begin{array}{cc}
  \mathbf{T}_{\frac{N}{2}}\left(\{\alpha_l\}_{l=\frac{3N}{2}}^{2N}\right)-i\mathbf{S}_{\frac{N}{2}}\left(\{\alpha_l\}_{l=N+1}^{\frac{3N}{2}}\right) & \mathbf{0} \\
  \mathbf{0} & \mathbf{T}_{\frac{N}{2}}\left(\{\alpha_l\}_{l=\frac{3N}{2}}^{2N}\right)+i\mathbf{S}_{\frac{N}{2}}\left(\{\alpha_l\}_{l=N+1}^{\frac{3N}{2}}\right) \\
\end{array}%
\right]\mathbf{\Pi}_N^H\;, \end{aligned}
\end{equation}
and
\begin{equation}\nonumber
 \left[\mathbf{\Pi}_N\right]_{ij}=\delta\left[2j-1-i\right]+\delta\left[2(j-\frac{N}{2})-i\right]\
\end{equation}
with $\delta[\cdot]$ denoting the delta function, giving $\delta[l]=1$ for $l=0$ and $\delta[l]=0$ for $l\neq 0$, and
$[\mathbf{\Pi}_N]_{ij}$ denotes the $(i,j)$-element of the $N \times N$ permutation matrix $\mathbf{\Pi}_N$.
\end{lemma}
\proof The proof is given in Appendix~\ref{sec:SVD_vonHTildeAllg}.\\
It is important to realize that $\mathbf{S}_N$ and $\mathbf{T}_N$ are constructed with different arguments.

The from $\mathbf{H}_{n_T}''$ in~(\ref{eq:Channel_VorZerleg}) in even and odd block structure re-sorted equivalent
channel matrix $\widetilde{\mathbf{H}}_{\frac{n_T}{2}}$ in~(\ref{eq:system_model_s1s3_colored_noise}) has exactly the
same structure as $\mathbf{M}_N$. Therefore, Lemma \ref{lem:FixedEigenv} can be directly applied to
$\widetilde{\mathbf{H}}_{\frac{n_T}{2}}$. To emphasis the usefullness of the resulting property of the QSTBC, we are
able to state the following theorem.
\begin{theorem}\label{theo:ConstEigenVec}
The left and right eigenvectors of the equivalent channel in~(\ref{eq:system_model_s1s3_colored_noise}) of QSTBC, which
fulfill the recursive construction rule of~(\ref{eq:neuer_Kanal}) are given by eq.~(\ref{eq:EigenvecVN}) and therefore
constant for any arbitrary channel realization.
\end{theorem}

\begin{remark}
Another important aspect of Lemma~\ref{lem:FixedEigenv} is the fact that the eigenvalues in $\mathbf{S}_N$ can be
obtained simply by adding the eigenvalues of $\mathbf{S}_\frac{N}{2}$ and $\mathbf{T}_\frac{N}{2}$ in an appropriate
manner as done in (\ref{eq:Building_Eigvalues}) (cf. Appendix~\ref{sec:SVD_vonHTildeAllg}), which will be used in the
following analysis of the QSTBC.
\end{remark}
\begin{lemma}\label{lem:EigenwerteMitMatrA}
Let $S$, $\mu_{n_T}^j$, $\nu_{n_T}^j$ be as in Lemma~\ref{lem:FixedEigenv}. The eigenvalues of the equivalent channel
matrix $\widetilde{\mathbf{H}}_{\frac{n_T}{2}}$ of the QSTBC are given by the recursive
equations~(\ref{eq:EigenvalSN}). Let $\mathbf{S}_{\frac{n_T}{2}}=\mathbf{D}_{\frac{n_T}{2}}\mathbf{D}_{\frac{n_T}{2}}$
and $(\tilde{\mu}_{n_T}^j)^2=\frac{2}{n_T}\mu_{n_T}^j$, where $\mu_{n_T}^j$, $1\leq j \leq \nicefrac{n_T}{2}$ are the
eigenvalues of $\mathbf{S}_{\frac{n_T}{2}}$. Then for any $n_T$ and $n_R$, the eigenvalues $(\tilde{\mu}_{n_T}^j)^2$ of
$\frac{2}{n_T}\mathbf{D}_{\frac{n_T}{2}}\mathbf{D}_{\frac{n_T}{2}}$ are obtained as follows
\begin{equation}\label{eq:eigenval_sum}
    (\tilde{\mu}_{n_T}^j)^2=\sum\limits_{i=1}^{n_R}\mathbf{h}_i^H\mathbf{A}_{n_T}^{j}\mathbf{h}_i,\; 1\leq j \leq
    \frac{n_T}{2}\;,
\end{equation}
where the matrices $\mathbf{A}_{n_T}^j$ with $j=1,3\dots,\nicefrac{n_T}{2}-1$ and $n_T=2^n$,$n\geq2$ are given as
\begin{equation}\label{eq:Def_MatrizenAnT}
    \mathbf{A}_{n_T}^{j}=\frac{1}{2}\left[\begin{array}{rr}
       \mathbf{A}_{\frac{n_T}{2}}^{j'}   & -\mathbf{B}_{\frac{n_T}{2}}^{j'} \\
      \mathbf{B}_{\frac{n_T}{2}}^{j'} & \mathbf{A}_{\frac{n_T}{2}}^{j'}     \\
    \end{array}\right],
\mathbf{A}_{n_T}^{j+1}=\frac{1}{2}\left[\begin{array}{rr}
       \mathbf{A}_{\frac{n_T}{2}}^{j'}   & \mathbf{B}_{\frac{n_T}{2}}^{j'} \\
      -\mathbf{B}_{\frac{n_T}{2}}^{j'} & \mathbf{A}_{\frac{n_T}{2}}^{j'}     \\
    \end{array}\right]
\end{equation}
with $\mathbf{B}_{n_T}^{j'}=i\mathbf{\Theta}_{n_T}\mathbf{A}_{n_T}^{j'}$ and ${j'}=\frac{j+1}{2}$.
\end{lemma}
\proof The proof is given in Appendix~\ref{sec:Proof_EigenwerteMitMatrA}.
\begin{theorem}[\cite{Muirhead}]\label{theo:Muirhead}
If $\mathbf{h}_i$ is $\mathcal{CN}(\mathbf{m}_i,\mathbf{I})$ and $\mathbf{P}$ is an $n_T \times n_T$ matrix then
$\mathbf{h}_i^H\mathbf{P}\mathbf{h}_i$ has a noncentral $\chi_k^2(\delta_{nc})$ distribution if and only if
$\mathbf{P}$ is idempotent ($\mathbf{P}^2=\mathbf{P}$), in which case the degrees of freedom is
$k=2\mathrm{rk}(\mathbf{P})=2\mathrm{tr}\{\mathbf{P}\}$ (where $\mathrm{rk}(\mathbf{P})$ and
$\mathrm{tr}\{\mathbf{P}\}$ denote the rank and trace of $\mathbf{P}$, respectively) and
$\delta_{nc}=\mathbf{m}_i^H\mathbf{P}\mathbf{m}_i$.
\end{theorem}
\begin{lemma}\label{lem:MatrAHermitIdempotent}
The matrices $\mathbf{A}_{n_T}^j$ are Hermitian and idempotent.
\end{lemma}
\proof The proof is given in Appendix~\ref{sec:Proof_MatrAHermitIdempotent}.\\
From Lemma~\ref{lem:EigenwerteMitMatrA} and~\ref{lem:MatrAHermitIdempotent}, it is now possible to prove the following
theorem.
\begin{theorem}\label{lem:unabh_Eigenwerte}
Let $\mathbf{D}_{\frac{n_T}{2}}\mathbf{D}_{\frac{n_T}{2}}$ defined as in Lemma~\ref{lem:EigenwerteMitMatrA} be the
diagonal eigenvalue matrix of the equivalent channel matrix of an QSTBC, which fulfill the recursive construction rule
of~(\ref{eq:neuer_Kanal}). Then for any $n_T$ and $n_R$, the eigenvalues $(\tilde{\mu}_{n_T}^j)^2$ of
$\frac{2}{n_T}\mathbf{D}_{\frac{n_T}{2}}\mathbf{D}_{\frac{n_T}{2}}$ are pairwise independent and identical noncentral
chi-square distributed with $4n_R$ degrees of freedom.
\end{theorem}
\begin{remark}
It is important to realize that $\mathbf{\Theta}_{n_T}^H=\mathbf{\Theta}_{n_T}$ and
$\mathbf{\Theta}_{n_T}^H\mathbf{A}_{n_T}^j\mathbf{\Theta}_{n_T}=\mathbf{A}_{n_T}^j$.
\end{remark}
\begin{remark}
Recall from section \ref{sec:Preprocessing}, that we have decoupled the system into two independent parts. The
derivation above holds therefore for both parts, i.e., each eigenvalue appears twice, once for each part.
\end{remark}
\proof The proof of Theorem~\ref{lem:unabh_Eigenwerte} is given in Appendix~\ref{sec:Proof_unabh_Eigenwerte}. 

\subsection{Noise pre-whitening}\label{sec:Noise pre-whitening}
Since $\widetilde{\mathbf{n}}$ in (\ref{eq:system_model_s1s3_colored_noise}) is colored noise, the next step is to
perform pre-whitening. With the knowledge of the theorems~\ref{theo:ConstEigenVec} and~\ref{lem:unabh_Eigenwerte} it is
easy to compute the pre-whitening filter $\mathbf{F}_{\mathrm{PW}}$ at the receiver now.
To this end, we need just the eigenvalue decomposition of $\widetilde{\mathbf{H}}_{\frac{n_T}{2}}$ given as
$\widetilde{\mathbf{H}}_{\frac{n_T}{2}}=\mathbf{V}_{\frac{n_T}{2}}\mathbf{S}_{\frac{n_T}{2}}\mathbf{V}_{\frac{n_T}{2}}^H
$ with $\mathbf{S}_{\frac{n_T}{2}}=\mathbf{D}_{\frac{n_T}{2}}\mathbf{D}_{\frac{n_T}{2}}$. Therefore the pre-whitening
filter is given as $\mathbf{F}_{\mathrm{PW}}=\mathbf{D}_{\frac{n_T}{2}}^{-1}\mathbf{V}_{\frac{n_T}{2}}^H $. By
multiplying $\mathbf{F}_{\mathrm{PW}}$ from the left to (\ref{eq:system_model_s1s3_colored_noise}) we arrive at
\begin{equation}\label{eq:system_model_nach_whiting}
    \widehat{\mathbf{y}}_{\mathrm{odd}} = \widehat{\mathbf{H}} \mathbf{x}_{\mathrm{odd}} + \mathbf{w}\;,
\end{equation}
 where the entries of $\mathbf{w}$ are mutually i.i.d. Gaussian processes again.

\begin{example} In the case of $n_T=4$ transmit antennas $\widehat{\mathbf{H}}$ in
(\ref{eq:system_model_nach_whiting}) is given as
\begin{equation}\label{eq:sys_model_nach_whiting_4MT}
   \widehat{\mathbf{H}} = \left[%
\begin{array}{rr}
  \tilde{\mu}_4^1    &     i\tilde{\mu}_4^1   \\
  \tilde{\mu}_4^2    &    -i\tilde{\mu}_4^2 \\
\end{array}%
\right]\;,
\end{equation}
 and
 \begin{equation}\label{eq:EigenwerteNt4}
 \tilde{\mu}_4^1 = \sqrt{\frac{\alpha_1+\alpha_2}{2}}, \text{ and }\; \tilde{\mu}_4^2 =
\sqrt{\frac{\alpha_1-\alpha_2}{2}}.
\end{equation}
\end{example}
\subsection{Linear maximum likelihood detection}\label{sec:trans_strat}
From theorems~\ref{theo:ConstEigenVec} and~\ref{lem:unabh_Eigenwerte}, it is now possible to determine
$\mathbf{\Gamma}$ adequately, resulting in a attractive system equation, which allows a very simple but effective
ML-decoding. To emphasize this property we formulate the following corollary.
\begin{corollary}\label{cor:LinMLDetec}
By choosing the matrix $\mathbf{\Gamma}$ in (\ref{eq:x_oddANDx_evenDEF}) as $\mathbf{\Gamma}=\mathbf{V}_{\frac{n_T}{2}}
$, (\ref{eq:system_model_nach_whiting}) can be rewritten as
\begin{equation}\label{eq:SystemWithD}
\widehat{\mathbf{y}}_{\mathrm{odd}} = \mathbf{D}_{\frac{n_T}{2}}\mathbf{s}^- + \mathbf{w}\;.
\end{equation}
At this point, the elements of $\mathbf{s}^-$ (and also $\mathbf{s}^+$) are completely decoupled, since they experience
no interference from each other. Thus, a linear ML-detector is able to detect the symbols (or elements) transmitted
from the antennas separately.
\end{corollary}
\begin{proof}
The matrix  $\widehat{\mathbf{H}}$ in (\ref{eq:system_model_nach_whiting}) can be decomposed as
\begin{equation}\label{eq:Zerlegung_H}
\widehat{\mathbf{H}}=\mathbf{D}_{\frac{n_T}{2}}\mathbf{V}_{\frac{n_T}{2}}^H\;,
\end{equation}
where $\mathbf{D}_{\frac{n_T}{2}}$ is a diagonal matrix containing the singular values of $\widehat{\mathbf{H}}$. Since
$\mathbf{V}_{\frac{n_T}{2}}$ is constant for all channel realizations, we can set
$\mathbf{\Gamma}=\mathbf{V}_{\frac{n_T}{2}}$ without any knowledge of the current channel realization. Using
(\ref{eq:Zerlegung_H}) in (\ref{eq:system_model_nach_whiting}) results in (\ref{eq:SystemWithD}). That concludes the
proof.
\end{proof}
\begin{example}
For $n_T=4$ transmit antennas, $\mathbf{V}_{2}$ and $\mathbf{D}_{2}$ are given as
\begin{equation}\nonumber
    \mathbf{V}_2=\frac{1}{\sqrt{2}}\left[\begin{array}{rr}
  1 & 1      \\
 -i & i \\
\end{array}\right], \mathbf{D}_2=\sqrt{2}\left[\begin{array}{rr}
  \tilde{\mu}_4^1 & 0      \\
  0 & \tilde{\mu}_4^2 \\
\end{array}\right]\;.
\end{equation}
\end{example}
\begin{example}
For $n_T=8$ transmit antennas, we have the following $\mathbf{V}_4$
\begin{equation}\nonumber
    \mathbf{V}_4=\frac{1}{\sqrt{4}}\left[\begin{array}{rrrr}
  1  &  1  &  1 &   1 \\
  -i & -i  &  i &   i \\
  i  & -i  & -i &   i \\
  1  & -1  &  1 &  -1 \\
\end{array}\right]
\end{equation}
and $\mathbf{D}_4=\sqrt{4}\mathrm{diag}(\tilde{\mu}_8^1,\dots,\tilde{\mu}_8^4)$ with
\begin{eqnarray}\label{eq:EigValNt8}
\begin{split}
\tilde{\mu}_8^1&=\sqrt{\nicefrac{(\alpha_1+\alpha_2+\alpha_3-\alpha_4)}{4}},\;
\tilde{\mu}_8^2=\sqrt{\nicefrac{(\alpha_1+\alpha_2-\alpha_3+\alpha_4)}{4}},\\
\tilde{\mu}_8^3&=\sqrt{\nicefrac{(\alpha_1-\alpha_2+\alpha_3+\alpha_4)}{4}}, \text{ and }
\tilde{\mu}_8^4=\sqrt{\nicefrac{(\alpha_1-\alpha_2-\alpha_3-\alpha_4)}{4}}.
\end{split}
\end{eqnarray}
\end{example}

\section{Performance analysis}\label{sec:PerfAnaly}
Based on these new insights, we provide some performance analysis in this section, where we focus on the case of
Rayleigh fading $(\mathbf{m}_i=\mathbf{0})$.
\subsection{Outage Probability $ P_{out}$ }\label{sec:outage} The mutual information of a MIMO system with $n_T$
transmit and $n_R$ receive antennas with no CSI at the transmitter and perfect CSI at the receiver by using the optimal
transmit strategy is given as\footnotetext[1]{In this paper, we use the same terminology as in~\cite{Telatar99}, i.e.
we use the term capacity only in the Shannon sense and distinguish therefore between the concept of outage mutual
information (OMI) and capacity.}~\cite{Telatar99}
\begin{equation}\nonumber
I = \log_2\det\left(\mathbf{I}_{n_R} +\frac{\rho}{n_T}\mathbf{H}\mathbf{H}^H\right)\;.
\end{equation}
The portion of the mutual information achieved  with QSTBC is
\begin{equation}
I_{Q} = \frac{2}{n_T}\log_2\det\left(\mathbf{I}_{\nicefrac{n_T}{2}}
+\frac{\rho}{n_T}\mathbf{D}_{\frac{n_T}{2}}\mathbf{D}_{\frac{n_T}{2}}\right)\label{eq:MutualInfoQSTBCdet}=
\frac{2}{n_T}\log_2\prod\limits_{j=1}^{\nicefrac{n_T}{2}}\left(1 +\frac{\rho\frac{n_T}{2}}{n_T}(\tilde{\mu}_{n_T}^j)^2
\right)\;.
\end{equation}
 The outage probability $P_{out}$ achievable with QSTBC is defined as the probability that $I_Q$ is smaller than a
 certain rate $R$, i.e.
\begin{equation}\nonumber
     P_{out}(R,n_T,n_R,\rho)=\mathrm{Pr}[I_Q<R]\;.
\end{equation}
Unfortunately, the exact analysis of $P_{out}$ is not available. Therefore, we provide a lower and upper bound in the
following.
\subsubsection{Lower bound}
\begin{proposition} The outage probability $P_{out}$ is lower bounded by
\begin{equation}
    P_{out}(R,n_T,n_R,\rho)\geq
  1-\exp\left(-\frac{n_T}{\rho}\left(2^R-1\right)\right)
        \sum\limits_{k=0}^{n_Tn_R-1}\frac{\left(\frac{n_T}{\rho}\left(2^R-1\right)\right)^k}{k!}
\;.\label{eq:PoutLowerBound}
\end{equation}
\end{proposition}
\begin{proof}
By using the arithmetic mean - geometric mean inequality, i.e.
\begin{equation}\nonumber
    \prod\limits_{l=1}^{L}a_l^{\nicefrac{1}{L}}\leq \frac{1}{L}\sum\limits_{l=1}^{L}a_l,\; a_l \geq 0
\end{equation}
with equality if and only if  $a_1=a_2=\dots=a_L$ we obtain an upper bound for $I_Q$ (and therefore a lower bound
on $P_{out}$) given as
\begin{equation}
    I_Q \leq \frac{2}{n_T}\log_2\left(\frac{2}{n_T}\left(\sum\limits_{j=1}^{\nicefrac{n_T}{2}}
                                    1 +\frac{\rho\frac{n_T}{2}}{n_T}(\tilde{\mu}_{n_T}^j)^2 \right)\right)^{\frac{n_T}{2}} \label{eq:MutualInfoUpperBound}
                                    =\log_2\left(1 +\frac{\rho}{n_T}\alpha_1 \right)=I_Q^u\;,
\end{equation}
where $\alpha_1 = \sum_{i=1}^{n_R}\sum_{j=1}^{n_T} |h_{ji}|^2$ for the general case of arbitrary $n_T$ similar to the
cases $n_T=4$ and $n_T=8$.
  The lower bound on the outage probability $P_{out}$ can be written as
  \begin{equation}
  P_{out}(R,n_T,n_R,\rho)=\mathrm{Pr}[I_Q<R]\geq\mathrm{Pr}[I_Q^u<R]=
  \mathrm{Pr}\left[\alpha_1<\frac{n_T}{\rho}\left(2^R-1\right)\right]\;. \nonumber
\end{equation}
Since $\alpha_1$ is chi-square distributed random variable with $2n_Tn_R$ degrees of freedom, $P_{out}$ is given
as~\cite[p.310,3.351(1)]{Gradshteyn} in (\ref{eq:PoutLowerBound}). That concludes the proof.
\end{proof}
\begin{corollary}\label{lem:Tightness_lowerBound}
The lower bound in (\ref{eq:PoutLowerBound}) gets tight for low
 SNR values or when $n_R$ increases.
 \end{corollary}
\proof The proof is given in Appendix~\ref{sec:Proof_TighnessLowBound}. 
%
\subsubsection{Upper bound} Using the positive definiteness of $\mathbf{D}_{\frac{n_T}{2}}\mathbf{D}_{\frac{n_T}{2}}$, (\ref{eq:MutualInfoQSTBCdet}) can be lower bounded as
\begin{equation}
I_{Q} \geq \frac{2}{n_T}\log_2\left(1
+\left(\frac{\rho}{n_T}\right)^{\frac{n_T}{2}}\det\left(\mathbf{D}_{\frac{n_T}{2}}\mathbf{D}_{\frac{n_T}{2}}\right)\right)=I_Q^l\;.\nonumber
\end{equation}
Thus, the upper bound on $P_{out}$ is given as
\begin{eqnarray}
    P_{out}(R,n_T,n_R,\rho)= \mathrm{Pr}[I_Q<R]
   \leq \mathrm{Pr}[I_Q^l<R]=P\Big[\det\left(\mathbf{D}_{\frac{n_T}{2}}\mathbf{D}_{\frac{n_T}{2}}\right)<\underbrace{\left(\frac{n_T}{\rho}\right)^{\nicefrac{n_T}{2}}\left(2^{\frac{Rn_T}{2}}-1\right)}_{\tilde{R}}\Big]
\;.\nonumber
\end{eqnarray}
For the special case of $n_T=4$ transmit antennas, $P_{out}$ is given as
\begin{equation}\nonumber
   P_{out}(R,n_T,n_R,\rho)\leq Pr[4(\tilde{\mu}_4^1)^2(\tilde{\mu}_4^2)^2<\tilde{R}]\;.
\end{equation}
Since $4(\tilde{\mu}_4^1)^2(\tilde{\mu}_4^2)^2$ is a product of two chi-square distributed random variables, both with
$m=4n_R$ degrees of freedom, $P_{out}$ is given by~\cite[p.365, eq.9.9.34]{Springer}
\begin{equation}\nonumber
P_{out}(R,m,\rho)\leq \int\limits_{0}^{\tilde{R}}\frac{y^{\frac{m}{2}-1}}{\Gamma(\frac{m}{2})^22^{m-1}}K_0(\sqrt{y})\;
dy
=\frac{{\tilde{R}}^{\frac{m}{2}}}{\Gamma(\frac{m}{2})^22^{m}}\left(\sum\limits_{k=0}^{\infty }{\frac {
\left(\ln(\frac{4}{\tilde{R}})+2\Psi(k+1)+\frac{1}{ \frac{m}{2}+k} \right) (\frac{\tilde{R}}{4})^k}
 { ( \frac{m}{2}+k)  (k!)^2}}\right)\;,\nonumber
\end{equation}
where $\Psi$ is the Psi function~\cite[p.943, eq.8.360]{Gradshteyn} and $\Gamma$ is the Gamma
function~\cite[p.XXXI]{Gradshteyn}. Note that for high SNR a useful and simple approximation of the outage probability
can be obtained by retaining only the first term (i.e. $k=0$) of the series within the upper bound.

Some simulation results of the performance of QSTBC and their interpretation are presented in the following section.
\subsection{Simulations}\label{seq:simulations} 
\begin{figure}[htbp]
     \centering
  \includegraphics[scale=0.75]{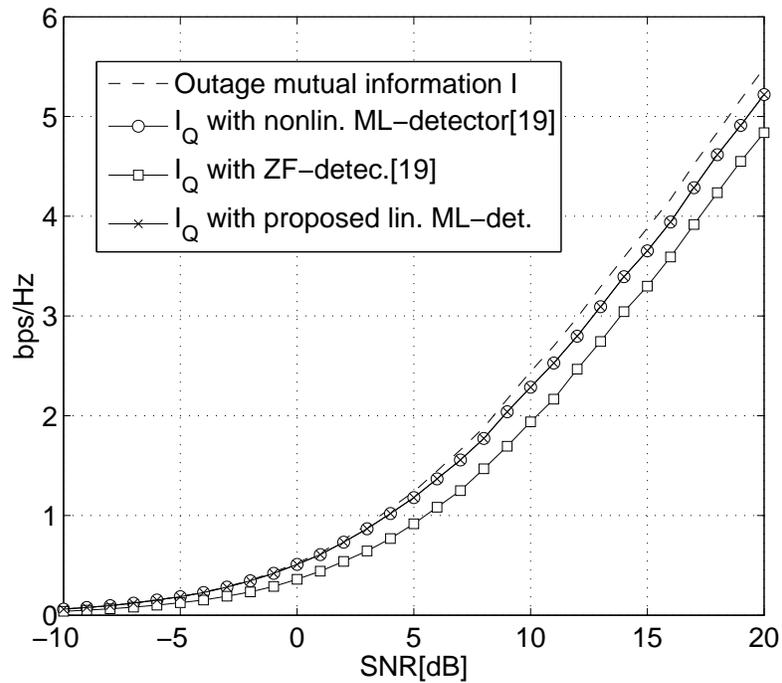}
  \caption{10\% Outage mutual information (OMI) of a MIMO system, our new approach,
  and the ML-and ZF-detector from~\cite{PapadiasFoschi01}
  with $n_T=4$ and $n_R= 1$.}
   \label{fig:capacity_nr1}
\end{figure}

In Fig.~\ref{fig:capacity_nr1}, the OMI of QSTBC $I_Q$ with our new
transmit strategy and our linear detector is compared with the nonlinear ML-detector and ZF-detector
in~\cite{PapadiasFoschi01}. Additionally, the OMI of a MIMO system with $n_T=4$ and $n_R=1$ is depicted. From the Fig.,
we observe that our new transmit strategy outperforms the ZF-detector of~\cite{PapadiasFoschi01} and achieve the same
portion of mutual information as the non-linear ML-detector presented in~\cite{PapadiasFoschi01}.

\begin{figure}[htbp]
     \centering
  \includegraphics[scale=0.6]{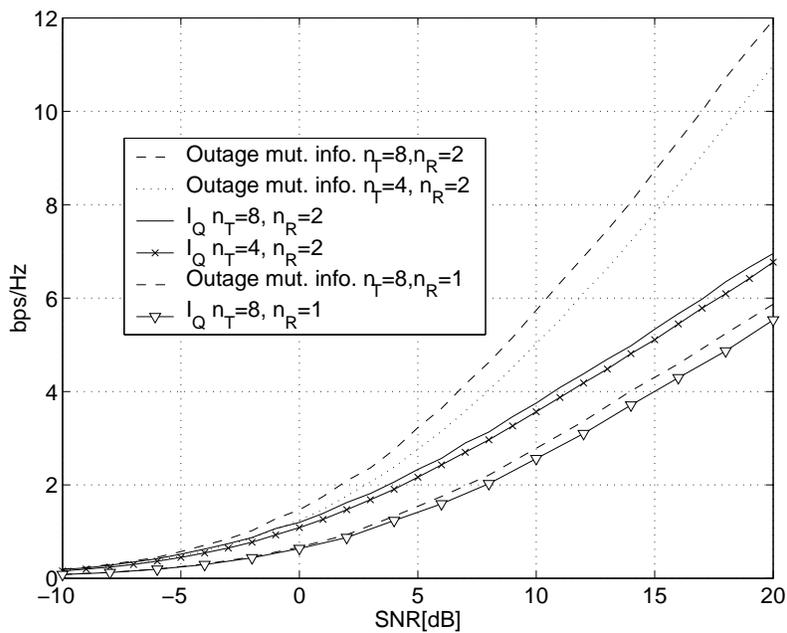}
  \caption{10\% Outage mutual information of a MIMO system  and our new approach
  with $n_T=4$,$n_T=8$ transmit and $n_R\geq 1$ receive antennas.}
   \label{fig:capacity_nr2}
\end{figure}

In Fig.~\ref{fig:capacity_nr2}, the performance of QSTBC in terms of OMI with $n_T=4$ and $n_T=8$ antennas is
depicted for $n_R\geq 1$.
For $n_R=1$, the performance with $n_T=8$ is similar to the case of $n_T=4$ transmit antennas (depicted in
Fig.~\ref{fig:capacity_nr2}), i.e. we achieve a significant fraction of the OMI. However, by increasing the number of
receive antennas, we observe in Fig.~\ref{fig:capacity_nr2}, that the performance of QSTBC with $n_T=8$ as well as with
$n_T=4$ is dramatically reduced in terms of achievable OMI.

\begin{figure}[htbp]
     \centering
  \includegraphics[scale=0.55]{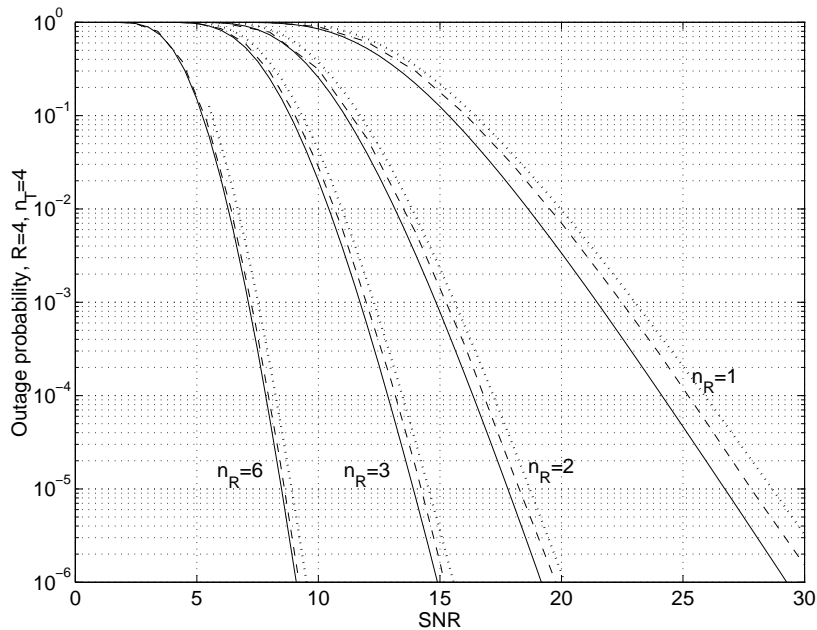}
  \caption{Outage probabilities of QSTBC (dashed lines), upper bound (dotted lines),  and lower bound(solid lines) for  $n_T=4$ transmit and different size of receive antennas $n_R$, Rate=4.}
   \label{fig:outage}
\end{figure}

\begin{figure}[htbp]
     \centering
  \includegraphics[scale=0.6]{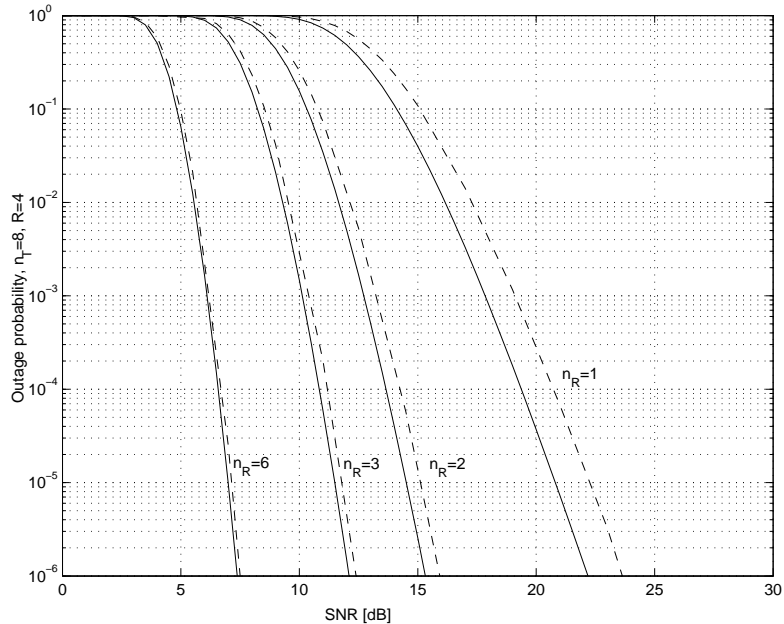}
  \caption{Outage probabilities of QSTBC (dashed lines) and lower bound(solid lines) for  $n_T=8$ transmit and different size of receive antennas $n_R$, Rate=4.}
   \label{fig:outage_nt8}
\end{figure}

In Fig.~\ref{fig:outage}, $P_{out}$ of QSTBC with $n_T=4$ transmit and $n_R=1$ to $n_R=3$ and $n_R=6$ receive antennas
is depicted. From the Fig. we observe that our lower bound on the performance of QSTBC with respect to $P_{out}$ gets
tight for increasing number of receive antennas. Even the upper bound performs very well and shows to be useful.
The performance of QSTBC with respect to $P_{out}$ is depicted in Fig.~\ref{fig:outage_nt8} for $n_T=8$ transmit
antennas. Similarly to the case of $n_T=4$ transmit antennas, the lower bound gets tight by using many antennas at
the receiver side.
\section{Conclusion}\label{seq:Conclusion}
In this paper, we generalized  the QSTBC in~\cite{PapadiasFoschi01} to $2^n$ transmit and an arbitrary number of
receive antennas. Our most important results are given in the theorems~\ref{theo:ConstEigenVec}
and~\ref{lem:unabh_Eigenwerte} in section~\ref{sec:EigValAndEigVecEquiChannel} and reveal the useful properties of the
resulting equivalent channel for QSTBC. In more details, theorem~\ref{theo:ConstEigenVec} state that the eigenvectors
of the equivalent channel are independent of each channel realization, i.e. they are fixed. This property can be
directly exploited for preprocessing, which allows a very efficient linear ML detection
(Corollary~\ref{cor:LinMLDetec}). In more details, we developed a new transmit strategy, which allows to apply a linear
ML-detector as in the case of OSTBC, such that the symbols from different antennas can be detected separately. The
performance of our linear detector in terms of mutual information is equal to the nonlinear ML-detector
in~\cite{PapadiasFoschi01}. The theorem~\ref{lem:unabh_Eigenwerte} offers insights in the statistics of QSTBC. It is
proved that the eigenvalues of the equivalent channel are i.i.d. noncentral $\chi_{4n_R}^2(\delta_{nc})$. These
insights were used to derive upper and lower bounds on the outage probability with QSTBC. It was shown, both
analytically and via simulations, that the lower bound gets tight for increasing number of receive antennas and also in
the low SNR-regime. From simulation results, we observed that the QSTBC approaches the outage mutual information only
in the case of $n_R=1$ receive antenna. By increasing the number of receive antennas, the loss in terms of mutual
information increases unbounded with the SNR.

\appendices

\section{Proof of Lemma \ref{lem:FixedEigenv}}\label{sec:SVD_vonHTildeAllg}
In the following, the arguments of $\mathbf{M}_N,\mathbf{N}_N,\mathbf{S}_N$ and $\mathbf{T}_N$ are omitted occasionally
in order to increase the readability of the paper.

\begin{proof}
The proof is done by the principle of induction. We start with the initial case $\mathbf{M}_4$.
From~(\ref{eq:StructureA_N}), it follows
\begin{align}
  \mathbf{M}_{4}=&\left[%
\begin{array}{rr}
  \mathbf{M}_{2}  & \mathbf{N}_{2} \\
  -\mathbf{N}_{2} & \mathbf{M}_{2} \\
\end{array}%
\right]\stackrel{(\ref{eq:SVD_A_2B_2})}{=}
  \left[%
\begin{array}{rr}
  \mathbf{V}_2\mathbf{S}_2\mathbf{V}_2^H  & \mathbf{V}_2\mathbf{T}_2\mathbf{V}_2^H \\
  -\mathbf{V}_2\mathbf{T}_2\mathbf{V}_2^H & \mathbf{V}_2\mathbf{S}_2\mathbf{V}_2^H \\
\end{array}%
\right] \nonumber \\
=&\left[%
\begin{array}{cc}
  \mathbf{V}_2 & \mathbf{0} \\
  \mathbf{0} & \mathbf{V}_2 \\
\end{array}%
\right]\left[%
\begin{array}{cc}
  \mathbf{S}_2 & \mathbf{T}_2 \\
  -\mathbf{T}_2 & \mathbf{S}_2 \\
\end{array}%
\right]\left[%
\begin{array}{cc}
  \mathbf{V}_2^H & \mathbf{0} \\
  \mathbf{0} & \mathbf{V}_2^H \\
\end{array}%
\right] \nonumber
\end{align}
\begin{align}
=&\left[%
\begin{array}{cc}
  \mathbf{V}_2 & \mathbf{0} \\
  \mathbf{0} & \mathbf{V}_2 \\
\end{array}%
\right]\mathbf{\Pi}_4
\left[%
\begin{array}{cccc}
  \mu_2^1  & \nu_2^1   & \multicolumn{2}{c}{\multirow{2}{*}{{\mbox{\huge $0$}}}} \\
  -\nu_2^1 & \mu_2^1   &         &  \\
  \multicolumn{2}{c}{\multirow{2}{*}{{\mbox{\huge $0$}}}}         & \mu_2^2   & \nu_2^2 \\
          &          & -\nu_2^2  & \mu_2^2 \\
\end{array}%
\right]\mathbf{\Pi}_4^H
\left[%
\begin{array}{cc}
  \mathbf{V}_2^H & \mathbf{0} \\
  \mathbf{0} & \mathbf{V}_2^H \\
\end{array}%
\right] \nonumber\\
=&\underbrace{\left[%
\begin{array}{cc}
  \mathbf{V}_2 & \mathbf{0} \\
  \mathbf{0} & \mathbf{V}_2 \\
\end{array}%
\right]\mathbf{\Pi}_4 \left[%
\begin{array}{cc}
  \mathbf{V}_2 & \mathbf{0} \\
  \mathbf{0} & \mathbf{V}_2 \\
\end{array}%
\right]}_{\mathbf{V}_4} \mathbf{S}_4
\underbrace{\left[%
\begin{array}{cc}
  \mathbf{V}_2^H & \mathbf{0} \\
  \mathbf{0} & \mathbf{V}_2^H \\
\end{array}%
\right]\mathbf{\Pi}_4^H
\left[%
\begin{array}{cc}
  \mathbf{V}_2^H & \mathbf{0} \\
  \mathbf{0} & \mathbf{V}_2^H \\
\end{array}%
\right]}_{\mathbf{V}_4^H}\;, \nonumber
\end{align}
with $\mathbf{\Pi}_4$ given as
\begin{equation}
\mathbf{\Pi}_4=
\left[%
\begin{array}{cccc}
  1 & 0 & 0 & 0 \\
  0 & 0 & 1 & 0 \\
  0 & 1 & 0 & 0 \\
  0 & 0 & 0 & 1 \\
\end{array}%
\right]\;,
\end{equation}
$ \mathbf{S}_4=\mathrm{diag}(\mu_4^1,\mu_4^2,\mu_4^3,\mu_4^4)$, where
\begin{equation}\label{eq:EigValNt4AusSumMuAndNu}
    \mu_4^1=\mu_2^1-i\nu_2^1\;,\;\mu_4^2=\mu_2^1+i\nu_2^1\;,\;\mu_4^3=\mu_2^2-i \nu_2^2\;,\text{ and }\;\mu_4^4=\mu_2^2
    +i\nu_2^2.
\end{equation}
which is equivalent to
\begin{equation}
    \mathbf{S}_4(\{\alpha_l\}_{l=1}^{4})=\mathbf{\Pi}_4\left[%
\begin{array}{cc}
  \mathbf{S}_2(\alpha_1,\alpha_2)-i\mathbf{T}_2(\alpha_3,\alpha_4) & \mathbf{0} \\
  \mathbf{0} & \mathbf{S}_2(\alpha_1,\alpha_2)+i\mathbf{T}_2(\alpha_3,\alpha_4) \\
\end{array}%
\right]\mathbf{\Pi}_4^H
\end{equation}
The same procedure applied to $\mathbf{N}_4$ results in a $\mathbf{T}_4$ given as
$\mathbf{T}_4=\mathrm{diag}(\nu_4^1,\nu_4^2,\dots,\nu_4^4)$ with
\begin{equation}
    \nu_4^1=\nu_2^3-i\mu_2^3\;,\;\nu_4^2=\nu_2^3+i\mu_2^3\;,\;\nu_4^3=\nu_2^4-i \mu_2^4\;,\text{ and }\;\nu_4^4=\nu_2^4
    +i\mu_2^4.
\end{equation}
which is equivalent to
\begin{equation}\nonumber
    \mathbf{T}_4(\{\alpha_l\}_{l=5}^{8})=\mathbf{\Pi}_4\left[%
\begin{array}{cc}
  \mathbf{T}_{2}(\alpha_7,\alpha_8)-i\mathbf{S}_{2}(\alpha_5,\alpha_6) & \mathbf{0} \\
  \mathbf{0} & \mathbf{T}_{2}(\alpha_7,\alpha_8)+i\mathbf{S}_{2}(\alpha_5,\alpha_6) \\
\end{array}%
\right]\mathbf{\Pi}_4^H\;.
\end{equation}
Now assume that the following hypothesis holds for $N=\nicefrac{K}{2}$, i.e.
\begin{equation}\label{eq:Induct_Hypothesis}
    \mathbf{M}_\frac{K}{2}=\mathbf{V}_\frac{K}{2}\mathbf{S}_\frac{K}{2}\mathbf{V}_\frac{K}{2}^H\; \text{ and } \;
    \mathbf{N}_\frac{K}{2}=\mathbf{V}_\frac{K}{2}\mathbf{T}_\frac{K}{2}\mathbf{V}_\frac{K}{2}^H\;,
\end{equation}
then the following inductive step is true
\begin{align}
  \mathbf{M}_{K} = & \left[%
\begin{array}{rr}
  \mathbf{M}_{\frac{K}{2}}  & \mathbf{N}_{\frac{K}{2}} \\
  -\mathbf{N}_{\frac{K}{2}} & \mathbf{M}_{\frac{K}{2}} \\
\end{array}%
\right]\stackrel{(\ref{eq:Induct_Hypothesis})}{=}
  \left[%
\begin{array}{rr}
  \mathbf{V}_\frac{K}{2}\mathbf{S}_\frac{K}{2}\mathbf{V}_\frac{K}{2}^H  & \mathbf{V}_\frac{K}{2}\mathbf{T}_\frac{K}{2}\mathbf{V}_\frac{K}{2}^H \\
  -\mathbf{V}_\frac{K}{2}\mathbf{T}_\frac{K}{2}\mathbf{V}_\frac{K}{2}^H & \mathbf{V}_\frac{K}{2}\mathbf{S}_\frac{K}{2}\mathbf{V}_\frac{K}{2}^H \\
\end{array}%
\right]\\
=&\left[%
\begin{array}{cc}
  \mathbf{V}_\frac{K}{2} & \mathbf{0} \\
  \mathbf{0} & \mathbf{V}_\frac{K}{2} \\
\end{array}%
\right]\left[%
\begin{array}{cc}
  \mathbf{S}_\frac{K}{2} & \mathbf{T}_\frac{K}{2} \\
  -\mathbf{T}_\frac{K}{2} & \mathbf{S}_\frac{K}{2} \\
\end{array}%
\right]\left[%
\begin{array}{cc}
  \mathbf{V}_\frac{K}{2}^H & \mathbf{0} \\
  \mathbf{0} & \mathbf{V}_\frac{K}{2}^H \\
\end{array}%
\right]\\
=&\left[%
\begin{array}{cc}
  \mathbf{V}_\frac{K}{2} & \mathbf{0} \\
  \mathbf{0} & \mathbf{V}_\frac{K}{2} \\
\end{array}%
\right]\mathbf{\Pi}_K
\left[%
\begin{array}{cccc}
  \mathbf{Q}_{\frac{K}{2}}^1 &                            &       \multicolumn{2}{c}{\multirow{2}{*}{{\mbox{\huge $0$}}}}  \\
                            & \mathbf{Q}_{\frac{K}{2}}^2  &    &       \\
                           \multicolumn{2}{c}{\multirow{2}{*}{{\mbox{\huge $0$}}}}   & \ddots &          \\
                       &         &       &\mathbf{Q}_{\frac{K}{2}}^{\frac{K}{2}} \\
\end{array}%
\right]\mathbf{\Pi}_K^H
\left[%
\begin{array}{cc}
  \mathbf{V}_\frac{K}{2}^H & \mathbf{0} \\
  \mathbf{0}\mathbf{} & \mathbf{V}_\frac{K}{2}^H \\
\end{array}%
\right],  \nonumber
\end{align}
\begin{align}
=&\underbrace{\left(\mathbf{I}_2 \otimes \mathbf{V}_\frac{K}{2} \right)\mathbf{\Pi}_K \left(\mathbf{I}_{\frac{K}{2}}
\otimes \mathbf{V}_2 \right)}_{\mathbf{V}_K} \mathbf{S}_K \underbrace{\left(\mathbf{I}_{\frac{K}{2}} \otimes
\mathbf{V}_2^H \right)\mathbf{\Pi}_K \left(\mathbf{I}_2 \otimes \mathbf{V}_{\frac{K}{2}}^H
\right)}_{\mathbf{V}_K^H}\nonumber
\end{align}
where
\begin{equation}\nonumber
 \mathbf{Q}_{\frac{K}{2}}^k =   \left[%
\begin{array}{cc}
  \mu_\frac{K}{2}^k & \nu_\frac{K}{2}^k \\
  -\nu_\frac{K}{2}^k & \mu_\frac{K}{2}^k \\
\end{array}%
\right]
\end{equation}
and $ \mathbf{S}_K=\mathrm{diag}(\mu_K^1,\mu_K^2,\dots,\mu_K^K)$ with
\begin{equation}\label{eq:Building_Eigvalues}
    \mu_K^{l-1}=\mu_\frac{K}{2}^{\frac{l}{2}}-i\nu_\frac{K}{2}^{\frac{l}{2}}\;\text{ and }\;\mu_K^{l}=\mu_\frac{K}{2}^{\frac{l}{2}}+i\nu_\frac{K}{2}^{\frac{l}{2}}
    \qquad \forall l=2,4,6,\dots,K
\end{equation}
which is equivalent to
\begin{equation}\nonumber
    \mathbf{S}_K=\mathbf{\Pi}_K\left[%
\begin{array}{cc}
  \mathbf{S}_{\frac{K}{2}}(\{\alpha_l\}_{l=1}^{\frac{K}{2}})-i\mathbf{T}_{\frac{K}{2}}(\{\alpha_l\}_{l=\frac{K}{2}+1}^{K}) & \mathbf{0} \\
  \mathbf{0} & \mathbf{S}_{\frac{K}{2}}(\{\alpha_l\}_{l=1}^{\frac{K}{2}})+i\mathbf{T}_{\frac{K}{2}}(\{\alpha_l\}_{l=\frac{K}{2}+1}^{K}) \\
\end{array}%
\right]\mathbf{\Pi}_K^H.
\end{equation}
The same procedure applied to $\mathbf{N}_K$ results in the same $\mathbf{V}_4$ and
$\mathbf{T}_K=\mathrm{diag}(\nu_K^1,\nu_K^2,\dots,\nu_K^K)$ with
\begin{equation}\label{eq:Build_EigvalNu}
    \nu_K^{l-1}=\nu_\frac{K}{2}^{\frac{K+l}{2}}-i\mu_\frac{K}{2}^{\frac{K+l}{2}}\;\text{ and }\;\nu_K^{l}=\nu_\frac{K}{2}^{\frac{K+l}{2}}+i\mu_\frac{K}{2}^{\frac{K+l}{2}}
    \qquad \forall l=2,4,6,\dots,K
\end{equation}
which is again equivalent to
\begin{equation}\nonumber
    \mathbf{T}_K=\mathbf{\Pi}_K\left[%
\begin{array}{cc}
  \mathbf{T}_{\frac{K}{2}}(\{\alpha_l\}_{l=K+\frac{K}{2}+1}^{2K})-i\mathbf{S}_{\frac{K}{2}}(\{\alpha_l\}_{l=K+1}^{K+\frac{K}{2}}) & \mathbf{0} \\
  \mathbf{0} & \mathbf{T}_{\frac{K}{2}}(\{\alpha_l\}_{l=K+\frac{K}{2}+1}^{2K})+i\mathbf{S}_{\frac{K}{2}}(\{\alpha_l\}_{l=K+1}^{K+\frac{K}{2}}) \\
\end{array}%
\right]\mathbf{\Pi}_K^H.
\end{equation}
Since the initial case of $N=4$ is true and the inductive step is true, the statement in (\ref{eq:SVD_Statement}) is
true for all $N=2^n$. That concludes the proof.
\end{proof}

\section{Proof of Lemma~\ref{lem:EigenwerteMitMatrA}}\label{sec:Proof_EigenwerteMitMatrA}
\begin{proof}
The proof is done by the principle of induction. The outline of the proof is as follows. For the initial case of
$n_T=4$, we need the eigenvalues $(\tilde{\mu}_2^1)^{2}$ and $(\tilde{\nu}_2^1)^{2}$ for $n_T=2$, i.e. the Alamouti
scheme, as indicated in~(\ref{eq:EigValNt4AusSumMuAndNu}). The first step is therefore to construct the eigenvalues for
$n_T=4$ with the eigenvalues of $n_T=2$. Using~(\ref{eq:Building_Eigvalues}) and~(\ref{eq:Build_EigvalNu}), we observe
that the eigenvalues for $n_T=4$ can be also obtained with the eigenvalues for $n_T=8$, which is the second step
revealing an important instruction of constructing eigenvalues $\mu_K$,$\nu_K$ from
$\mu_\frac{K}{2}$,$\nu_\frac{K}{2}$. It follows the hypothesis and the inductive step concluding the proof.

Now, we start with the well known Alamouti scheme. By applying the Alamouti scheme ($n_T=2$),
$\frac{2}{n_T}\mathbf{D}_{\frac{n_T}{2}}\mathbf{D}_{\frac{n_T}{2}}$ as well as $\mathbf{x}_{\mathrm{odd}}$ and
$\mathbf{x}_{\mathrm{even}}$ are only scalars. Thus, the only eigenvalue of $\mathbf{D}_{1}\mathbf{D}_{1}$ of the
decomposed system model for the part with $\mathbf{x}_{\mathrm{odd}}$ (and similar for $\mathbf{x}_{\mathrm{even}}$) is
given as
\begin{equation}\label{eq:EigenValAlam}
(\tilde{\mu}_2^1)^{2}=\sum_{i=1}^{n_R}\alpha_1(h_{1i},h_{2i})=\sum_{i=1}^{n_R}\sum_{j=1}^{n_T=2}|h_{ji}|^2=\sum_{i=1}^{n_R}\mathbf{h}_{1\rightarrow
2,i}^H\mathbf{A}_2^1\mathbf{h}_{1\rightarrow 2,i}
\end{equation}
where $\mathbf{h}_{k\rightarrow l,i}=[h_{ki},\dots,h_{li}]^T$ and
\begin{equation}
\mathbf{A}_2^1=\mathbf{A}_2=\mathbf{I}_2
\end{equation}
Similarly,
\begin{equation}\label{eq:EigenValAlamNu}
(\tilde{\nu}_2^1)^{2}=\sum_{i=1}^{n_R}\alpha_2(h_{3i},h_{4i})=\sum_{i=1}^{n_R}\mathbf{h}_{3\rightarrow
4,i}^H\mathbf{A}_2^1\mathbf{h}_{3\rightarrow 4,i}
\end{equation}

 We are now able to start with the initial case of $n_T=4$.
 The first eigenvalue of the QSTBC for the part with $\mathbf{x}_{\mathrm{odd}}$ (and
similar for $\mathbf{x}_{\mathrm{even}}$)  is given as in~(\ref{eq:Building_Eigvalues}) (with
$(\tilde{\mu}_{n_T}^j)^2=\frac{2}{n_T}\mu_{n_T}^j$ and $(\tilde{\nu}_{n_T}^j)^2=\frac{2}{n_T}\nu_{n_T}^j$)
\begin{align}
(\tilde{\mu}_4^1)^2&=(\tilde{\mu}_4^1(h_{1i},\dots,h_{4i}))^2= \tilde{\mu}_2^1-i\tilde{\nu}_2^1\stackrel{(\ref{eq:EigenwerteNt4})}{=}\sum_{i=1}^{n_R}\frac{1}{2}(\alpha_1(h_{1i},\dots,h_{4i}) + \alpha_2(h_{1i},\dots,h_{4i})) \\
&\stackrel{(\ref{eq:alpha1u2forNt4})}{=} \sum_{i=1}^{n_R}  \frac{1}{2}(\sum_{j=1}^{n_T} |h_{ji}|^2 +
2\mathrm{Im}(h_{1i}^*h_{3i}+h_{4i}^*h_{2i})) \nonumber
\end{align}
\begin{align}
\phantom{(\tilde{\mu}_4^1)^2}&=\sum_{i=1}^{n_R}\frac{1}{2}\left([\mathbf{h}_{1\rightarrow 2,i}^H \mathbf{h}_{3\rightarrow 4,i}^H]\left[%
\begin{array}{c}
  \mathbf{h}_{1\rightarrow 2,i} \\
  \mathbf{h}_{3\rightarrow 4,i} \\
\end{array}%
\right]- i\left[%
\begin{array}{cccc}
  -h_{3i}^*& h_{4i}^* & h_{1i}^* & -h_{2i}^* \\
\end{array}%
\right]\left[%
\begin{array}{c}
  \mathbf{h}_{1\rightarrow 2,i} \\
  \mathbf{h}_{3\rightarrow 4,i} \\
\end{array}%
\right]\right)\nonumber\end{align}
\begin{align}
\phantom{(\tilde{\mu}_4^1)^2}&= \sum_{i=1}^{n_R}\mathbf{h}_{1\rightarrow 4,i}^H\frac{1}{2}\left[%
\begin{array}{cc}
  \mathbf{A}_2 & \mathbf{0} \\
  \mathbf{0} & \mathbf{A}_2 \\
\end{array}%
\right]\mathbf{h}_{1\rightarrow 4,i}-i\left[%
\begin{array}{cccc}
   h_{1i}^* & -h_{2i}^* & -h_{3i}^*& h_{4i}^*\\
\end{array}%
\right]\frac{1}{2}\left[%
\begin{array}{cc}
  \mathbf{0} & \mathbf{A}_2 \\
   \mathbf{A}_2 & \mathbf{0} \\
\end{array}%
\right]\mathbf{h}_{1\rightarrow 4,i} \label{eq:ExpansNt2ToNt4}\\
&=\sum_{i=1}^{n_R}\mathbf{h}_{1\rightarrow 4,i}^H\frac{1}{2}\left[%
\begin{array}{cc}
  \mathbf{A}_2 & \mathbf{0} \\
  \mathbf{0} & \mathbf{A}_2 \\
\end{array}%
\right]\mathbf{h}_{1\rightarrow 4,i}-i\mathbf{h}_{1\rightarrow 4,i}^H\frac{1}{2}\left[%
\begin{array}{cc}
  \mathbf{\Theta}_2 & \mathbf{0} \\
  \mathbf{0} & -\mathbf{\Theta}_2 \\
\end{array}%
\right]\left[%
\begin{array}{cc}
  \mathbf{0} & \mathbf{A}_2 \\
   \mathbf{A}_2 & \mathbf{0} \\
\end{array}%
\right]
  \mathbf{h}_{1\rightarrow 4,i}\nonumber\\
 & =\sum_{i=1}^{n_R}\mathbf{h}_{1\rightarrow 4,i}^H\frac{1}{2}\left(\left[%
\begin{array}{cc}
  \mathbf{A}_2 & \mathbf{0} \\
  \mathbf{0} & \mathbf{A}_2 \\
\end{array}%
\right]-\left[%
\begin{array}{cc}
  \mathbf{0} & i\mathbf{\Theta}_2\mathbf{A}_2 \\
   -i\mathbf{\Theta}_2\mathbf{A}_2 & \mathbf{0} \\
\end{array}%
\right]\right)\mathbf{h}_{1\rightarrow 4,i} =\mathbf{h}_{1\rightarrow 4,i}^H\mathbf{A}_4^1\mathbf{h}_{1\rightarrow
4,i}\;,\label{eq:EigValNt4FromAlpha}
\end{align}
where
\begin{equation}\nonumber
\mathbf{A}_4^1=\frac{1}{2}\left[%
\begin{array}{cc}
  \mathbf{A}_2 & -i\mathbf{\Theta}_2\mathbf{A}_2 \\
  i\mathbf{\Theta_2}\mathbf{A}_2 & \mathbf{A}_2 \\
\end{array}%
\right]\;.
\end{equation}
In an analogous manner, we get $(\tilde{\nu}_4^1)^2$ given as
\begin{equation}\nonumber
(\tilde{\nu}_4^1)^2=\mathbf{h}_{5\rightarrow 8,i}^H\mathbf{A}_4^1\mathbf{h}_{5\rightarrow 8,i}
\end{equation}

On the other hand, with~(\ref{eq:Building_Eigvalues}) we have
\begin{align}
(\tilde{\mu}_4^1)^2&=(\tilde{\mu}_4^1(h_{1i},\dots,h_{8i}))^2
=\sum_{i=1}^{n_R}\frac{1}{2}((\tilde{\mu}_8^1)^2+(\tilde{\mu}_8^2)^2)
\stackrel{(\ref{eq:EigValNt8})}{=}\sum_{i=1}^{n_R}\frac{1}{2}\frac{\alpha_1(h_{1i},\dots,h_{8i})+\alpha_2(h_{1i},\dots,h_{8i})}{2}\nonumber\\
&\stackrel{(\ref{eq:alpha1bis4Nt8})}{=}\sum_{i=1}^{n_R}\frac{1}{4}\left(\sum_{j=1}^{8} |h_{ji}|^2+
2\mathrm{Im}(h_{1i}^*h_{3i}+h_{4i}^*h_{2i}+h_{5i}^*h_{7i}+h_{8i}^*h_{6i})\right)\nonumber\\
&=\sum_{i=1}^{n_R}\frac{1}{4}\left(\mathbf{h}_{1\rightarrow 8,i}^H
  \mathbf{h}_{1\rightarrow 8,i}- i\left[%
\begin{array}{cccccccc}
  -h_{3i}^*& h_{4i}^* & h_{1i}^* & -h_{2i}^* &-h_{7i}^*& h_{8i}^* & h_{5i}^* & -h_{6i}^* \\
\end{array}%
\right]
  \mathbf{h}_{1\rightarrow 8,i}\right)\nonumber\\
&=\sum_{i=1}^{n_R}\frac{1}{4}\mathbf{h}_{1\rightarrow 8,i}^H
  \mathbf{h}_{1\rightarrow 8,i}-i\mathbf{h}_{1\rightarrow 8,i}^H\frac{1}{4}\left[%
\begin{array}{cccc}
  \mathbf{\Theta}_2 &  & \multicolumn{2}{c}{\multirow{2}{*}{{\mbox{\huge $0$}}}} \\
   & -\mathbf{\Theta}_2 & & \\
  \multicolumn{2}{c}{\multirow{2}{*}{{\mbox{\huge $0$}}}} & \mathbf{\Theta}_2 &  \\
 & &  & -\mathbf{\Theta}_2 \\
\end{array}%
\right]\left[%
\begin{array}{cccc}
  \mathbf{0} & \mathbf{A}_2 &\multicolumn{2}{c}{\multirow{2}{*}{{\mbox{\huge $0$}}}} \\
   \mathbf{A}_2 & \mathbf{0} & &\\
 \multicolumn{2}{c}{\multirow{2}{*}{{\mbox{\huge $0$}}}} & \mathbf{0} & \mathbf{A}_2 \\
 & &  \mathbf{A}_2 & \mathbf{0} \\
\end{array}%
\right]
  \mathbf{h}_{1\rightarrow 8,i}\nonumber
  \end{align}
\begin{align}
\phantom{(\tilde{\mu}_4^1)^2}
 &=\sum_{i=1}^{n_R}\frac{1}{4}\mathbf{h}_{1\rightarrow 8,i}^H\left(\left[%
\begin{array}{cccc}
  \mathbf{A}_2 &  & \multicolumn{2}{c}{\multirow{2}{*}{{\mbox{\huge $0$}}}} \\
   & \mathbf{A}_2 & & \\
  \multicolumn{2}{c}{\multirow{2}{*}{{\mbox{\huge $0$}}}} & \mathbf{A}_2 &  \\
 & &  & \mathbf{A}_2 \\
\end{array}%
\right]
  -i\left[%
\begin{array}{cccc}
  \mathbf{0} & \mathbf{\Theta}_2\mathbf{A}_2 &\multicolumn{2}{c}{\multirow{2}{*}{{\mbox{\huge $0$}}}} \\
   -\mathbf{\Theta}_2\mathbf{A}_2 &\mathbf{0}& &\\
 \multicolumn{2}{c}{\multirow{2}{*}{{\mbox{\huge $0$}}}} & \mathbf{0} & \mathbf{\Theta}_2\mathbf{A}_2 \\
 & &  -\mathbf{\Theta}_2\mathbf{A}_2 & \mathbf{0} \\
\end{array}%
\right]\right)
  \mathbf{h}_{1\rightarrow 8,i}\nonumber
  \end{align}
  \begin{align}
 \phantom{(\tilde{\mu}_4^1)^2}
  &=\sum_{i=1}^{n_R}\frac{1}{4}\mathbf{h}_{1\rightarrow 8,i}^H\left(
  \left[%
\begin{array}{cccc}
  \mathbf{A}_2 & -i\mathbf{\Theta}_2\mathbf{A}_2 &\multicolumn{2}{c}{\multirow{2}{*}{{\mbox{\huge $0$}}}} \\
   i\mathbf{\Theta}_2\mathbf{A}_2 & \mathbf{A}_2 & &\\
 \multicolumn{2}{c}{\multirow{2}{*}{{\mbox{\huge $0$}}}} & \mathbf{A}_2 & -i\mathbf{\Theta}_2\mathbf{A}_2 \\
 & &  i\mathbf{\Theta}_2\mathbf{A}_2 & \mathbf{A}_2 \\
\end{array}%
\right]\right)
  \mathbf{h}_{1\rightarrow 8,i}\nonumber\\
    &=\sum_{i=1}^{n_R}\mathbf{h}_{1\rightarrow 8,i}^H\left(\frac{1}{2}
  \left[%
\begin{array}{cc}
  \mathbf{A}_4^1 & \mathbf{0} \\
   \mathbf{0} & \mathbf{A}_4^1 \\
\end{array}%
\right]\right)
  \mathbf{h}_{1\rightarrow 8,i}=\sum_{i=1}^{n_R}\mathbf{h}_{1\rightarrow 8,i}^H\frac{1}{2}\left(
  \mathbf{I}_2 \otimes \mathbf{A}_4^1   \right)
  \mathbf{h}_{1\rightarrow 8,i}\label{eq:EigValNt4KronProd}
   \end{align}
  \begin{align}
 \phantom{(\tilde{\mu}_4^1)^2}
&=\sum_{i=1}^{n_R}\frac{1}{2}[\mathbf{h}_{1\rightarrow 4,i}^H \mathbf{h}_{5\rightarrow 8,i}^H]\left[%
\begin{array}{cc}
  \mathbf{A}_4^1 & \mathbf{0} \\
   \mathbf{0} & \mathbf{A}_4^1 \\
\end{array}%
\right]\left[%
\begin{array}{c}
  \mathbf{h}_{1\rightarrow 4,i} \\
  \mathbf{h}_{5\rightarrow 8,i} \\
\end{array}%
\right]=\sum_{i=1}^{n_R}\frac{1}{2}\mathbf{h}_{1\rightarrow 4,i}^H\mathbf{A}_4^1\mathbf{h}_{1\rightarrow
4,i}+\frac{1}{2}\mathbf{h}_{5\rightarrow 8,i}^H\mathbf{A}_4^1\mathbf{h}_{5\rightarrow 8,i}\label{eq:EigValNt4FromNt8}
\end{align}
In an analogous manner, we get $(\tilde{\nu}_4^1)^2$ given as
\begin{equation}\nonumber
(\tilde{\nu}_4^1)^2=\sum_{i=1}^{n_R}\frac{1}{2}[\mathbf{h}_{1\rightarrow 4,i}^H \mathbf{h}_{5\rightarrow 8,i}^H]\left[%
\begin{array}{cc}
   \mathbf{0} &i\mathbf{\Theta}_k\mathbf{A}_4^1  \\
    -i\mathbf{\Theta}_k\mathbf{A}_4^1 &\mathbf{0}  \\
\end{array}%
\right]\left[%
\begin{array}{c}
  \mathbf{h}_{1\rightarrow 4,i} \\
  \mathbf{h}_{5\rightarrow 8,i} \\
\end{array}%
\right]
\end{equation}
Since the eigenvalues $(\tilde{\mu}_4^2)^2$ and $(\tilde{\nu}_4^2)^2$ can be obtained very easily in a similar way, we
omit the derivations here.

Comparing~(\ref{eq:EigValNt4FromNt8}) with~(\ref{eq:EigValNt4FromAlpha}) shows that in order to get the eigenvalues of
$n_T=8$, the eigenvalues of $n_T=4$ have to be expanded by using the Kronecker product of $\mathbf{I}_2$ and
$\mathbf{A}_4$ , divided by $\frac{1}{2}$ in order to incorporate the channel entries $h_{5i},\dots,h_{8i}$ as given
in~(\ref{eq:EigValNt4KronProd}). Actually, this can also be observed in the expansion from $n_T=2$ to $n_T=4$ by
comparing~(\ref{eq:EigenValAlam}) with the first addend in~(\ref{eq:ExpansNt2ToNt4}).

Now assume that the following hypothesis holds
\begin{eqnarray}\label{eq:EigValNtSubToh1Tohk}
(\tilde{\mu}_{k}^j(h_{1i},\dots,h_{ki}))^2 &=& \sum_{i=1}^{n_R}\mathbf{h}_{1\rightarrow k,i}^H\mathbf{A}_k^j\mathbf{h}_{1\rightarrow k,i} \\
(\tilde{\mu}_{k}^j(h_{1i},\dots,h_{2ki}))^2 &=& \sum_{i=1}^{n_R}\mathbf{h}_{1\rightarrow 2k,i}^H\frac{1}{2}\left[%
\begin{array}{cc}
  \mathbf{A}_k^j & \mathbf{0} \\
  \mathbf{0} & \mathbf{A}_k^j \\
\end{array}%
\right]\mathbf{h}_{1\rightarrow 2k,i}\label{eq:EigValNtSubToh1Toh2k}\;,
\end{eqnarray}
Similarly
\begin{eqnarray}\nonumber
(\tilde{\nu}_{k}^j(h_{1i},\dots,h_{ki}))^2 &=& \sum_{i=1}^{n_R}\mathbf{h}_{k+1\rightarrow 2k,i}^H\mathbf{A}_k^j\mathbf{h}_{k+1\rightarrow 2k,i} \\
(\tilde{\nu}_{k}^j(h_{1i},\dots,h_{2ki}))^2 &=& \sum_{i=1}^{n_R}\mathbf{h}_{1\rightarrow 2k,i}^H\frac{1}{2}\left[%
\begin{array}{cc}
  \mathbf{0}& i\mathbf{\Theta}_k\mathbf{A}_k^j  \\
  -i\mathbf{\Theta}_k\mathbf{A}_k^j & \mathbf{0} \\
\end{array}%
\right]\mathbf{h}_{1\rightarrow 2k,i}\label{eq:EigValnuNtSubToh1Toh2k}\;,
\end{eqnarray}
then the following inductive step is true
\begin{align}
(\tilde{\mu}_{2k}^{j,j+1})^2\stackrel{\phantom{cc}(\ref{eq:Building_Eigvalues})\phantom{cc}}{=}
&\sum_{i=1}^{n_R}(\tilde{\mu}_{k}^{j'}(h_{1i},\dots,h_{2ki}))^2\mp(\tilde{\nu}_{k}^{j'}(h_{1i},\dots,h_{2ki}))^2 \nonumber\\
\stackrel{(\ref{eq:EigValNtSubToh1Toh2k}),(\ref{eq:EigValnuNtSubToh1Toh2k})}{=}
&\sum_{i=1}^{n_R}[\mathbf{h}_{1\rightarrow k,i}^H \mathbf{h}_{k+1\rightarrow 2k,i}^H]\frac{1}{2}\left[%
\begin{array}{cc}
  \mathbf{A}_k^{j'} & \mathbf{0} \\
  \mathbf{0} & \mathbf{A}_k^{j'} \\
\end{array}%
\right]\left[%
\begin{array}{c}
  \mathbf{h}_{1\rightarrow k,i} \\
  \mathbf{h}_{k+1\rightarrow 2k,i} \\
\end{array}%
\right]\\
\phantom{=}& \mp [\mathbf{h}_{1\rightarrow k,i}^H \mathbf{h}_{k+1\rightarrow 2k,i}^H]\frac{1}{2}\left[%
\begin{array}{cc}
  \mathbf{0} & i\Theta_k\mathbf{A}_k^{j'} \\
  -i\Theta_k\mathbf{A}_k^{j'} & \mathbf{0} \\
\end{array}%
\right]\left[%
\begin{array}{c}
  \mathbf{h}_{1\rightarrow k,i} \\
  \mathbf{h}_{k+1\rightarrow 2k,i} \\
\end{array}%
\right]\nonumber
\end{align}
\begin{align}
\phantom{(\tilde{\mu}_{2k}^{j,j+1})^2}\stackrel{\phantom{(\ref{eq:EigValNtSubToh1Toh2k}),(\ref{eq:EigValnuNtSubToh1Toh2k})}}{=}
&\sum_{i=1}^{n_R}[\mathbf{h}_{1\rightarrow k,i}^H \mathbf{h}_{k+1\rightarrow 2k,i}^H]\frac{1}{2}\left[%
\begin{array}{cc}
  \mathbf{A}_k^{j'} & \mp i\Theta_k\mathbf{A}_k^{j'} \\
  \pm i\Theta_k\mathbf{A}_k^{j'} & \mathbf{A}_k^{j'} \\
\end{array}%
\right]\left[%
\begin{array}{c}
  \mathbf{h}_{1\rightarrow k,i} \\
  \mathbf{h}_{k+1\rightarrow 2k,i} \\
\end{array}%
\right] \nonumber
\end{align}
\begin{align}
\phantom{(\tilde{\mu}_{2k}^{j,j+1})^2}\stackrel{\phantom{(\ref{eq:EigValNtSubToh1Toh2k}),(\ref{eq:EigValnuNtSubToh1Toh2k})}}{=}
 &\sum_{i=1}^{n_R}\mathbf{h}_{1\rightarrow 2k,i}^H\mathbf{A}_{2k}^{j,j+1}\mathbf{h}_{1\rightarrow 2k,i}=\sum_{i=1}^{n_R}
\mathbf{h}_{1\rightarrow 2k,i}^H\frac{1}{2}\left[%
\begin{array}{cc}
  \mathbf{A}_k^{j'} & \mp i\Theta_k\mathbf{A}_k^{j'} \\
  \pm i\Theta_k\mathbf{A}_k^{j'} & \mathbf{A}_k^{j'} \\
\end{array}%
\right]\mathbf{h}_{1\rightarrow 2k,i}\;,
\end{align}
with $j=1,3,\dots,\nicefrac{n_T}{2}-1$ and ${j'}=\frac{j+1}{2}$.
That concludes the proof.
\end{proof}
\section{Proof of Lemma~\ref{lem:MatrAHermitIdempotent}}\label{sec:Proof_MatrAHermitIdempotent}
\begin{proof}
With~\ref{eq:Def_MatrizenAnT}, $(\mathbf{A}_{n_T}^{j})^H$ and $(\mathbf{A}_{n_T}^{j+1})^H$,
$j=1,3,\dots,\nicefrac{n_T}{2}$ and ${j'}=\frac{j+1}{2}$ are given as
\begin{equation}\nonumber
    (\mathbf{A}_{n_T}^{j,j+1})^H=\frac{1}{2}\left[\begin{array}{cc}
       (\mathbf{A}_{\frac{n_T}{2}}^{j'})^H   & \mp i\mathbf{\Theta}_{n_T}(\mathbf{A}_{n_T}^{j'})^H \\
      \pm i\mathbf{\Theta}_{n_T}(\mathbf{A}_{n_T}^{j'})^H & (\mathbf{A}_{\frac{n_T}{2}}^{j'})^H     \\
    \end{array}\right]\;.
    \end{equation}
Thus, $\mathbf{A}_{n_T}^{j,j+1}$ are only Hermitian, if $\mathbf{A}_{\frac{n_T}{2}}^{j'}$ is Hermitian. Since
$\mathbf{A}_{2}$ is Hermitian, it follows that $\mathbf{A}_{n_T}^j$, $n_T=2^n$,$j=1,3,\dots,\nicefrac{n_T}{2}$ are
Hermitian. Similarly,
\begin{equation}
    (\mathbf{A}_{n_T}^{j,j+1})^H\mathbf{A}_{n_T}^{j,j+1}=\mathbf{A}_{n_T}^{j,j+1}\mathbf{A}_{n_T}^{j,j+1}=\frac{1}{4}\left[\begin{array}{cc}
       2\mathbf{A}_{\frac{n_T}{2}}^{j'}\mathbf{A}_{\frac{n_T}{2}}^{j'}   & \mp i2\mathbf{\Theta}_{n_T}\mathbf{A}_{n_T}^{j'}\mathbf{A}_{n_T}^{j'} \\
      \pm i2\mathbf{\Theta}_{n_T}\mathbf{A}_{n_T}^{j'}\mathbf{A}_{n_T}^{j'} & 2\mathbf{A}_{\frac{n_T}{2}}^{j'}\mathbf{A}_{\frac{n_T}{2}}^{j'}    \\
    \end{array}\right]\;.
    \end{equation}
  Thus, $\mathbf{A}_{n_T}^{j,j+1}$ are only idempotent, if $\mathbf{A}_{\frac{n_T}{2}}^{j'}$ is idempotent. Since
$\mathbf{A}_{2}$ is idempotent, it follows that $\mathbf{A}_{n_T}^j$, $n_T=2^n$,$j=1,3,\dots,\nicefrac{n_T}{2}$ are
idempotent. That concludes the proof.
\end{proof}
\section{Proof of theorem~\ref{lem:unabh_Eigenwerte}}\label{sec:Proof_unabh_Eigenwerte}
\begin{proof}
We first prove the independency of the eigenvalues. Since the matrices $\mathbf{A}_{n_T}^{(j)}$ are Hermitian and
idempotent, we are able to rewrite~(\ref{eq:eigenval_sum}) as
\begin{equation}\nonumber
    (\mu_j^{n_T})^2=\sum\limits_{i=1}^{n_R}\mathbf{h}_i^H(\mathbf{A}_{n_T}^{j})^H\mathbf{A}_{n_T}^{j}\mathbf{h}_i
    =\sum\limits_{i=1}^{n_R}\|\mathbf{A}_{n_T}^{j}\mathbf{h}_i\|^2 \;.
\end{equation}
Independency between the eigenvalues in the Gaussian case is given if and only if the eigenvalues are uncorrelated,
i.e.
\begin{equation}
    E[(\mathbf{A}_{n_T}^{j}\mathbf{h}_i)^H\mathbf{A}_{n_T}^{k}\mathbf{h}_i]=0\; \text{    $\forall j, j\neq
    k$}\;, \nonumber
\end{equation}
which is fulfilled if
\begin{equation}\label{eq:Unabh_Eigenwerte}
    (\mathbf{A}_{n_T}^{j})^H\mathbf{A}_{n_T}^{k}=\mathbf{A}_{n_T}^{j}\mathbf{A}_{n_T}^{k}=\mathbf{0}\;
    \text{    $\forall j, j\neq
    k$}\;.
\end{equation}
By applying the eigenvalue decomposition to (\ref{eq:Unabh_Eigenwerte}), one has to distinguish between the case,
where the eigenvalues are given as
\begin{equation}\label{eq:EigenValuesI}
    (\mathbf{I}-\mathbf{\Theta}_{n_T})\mathbf{A}_{\frac{n_T}{2}}^{j'}(\mathbf{I}+\mathbf{\Theta}_{n_T})\mathbf{A}_{\frac{n_T}{2}}^{k'},\;(\mathbf{I}+\mathbf{\Theta}_{n_T})\mathbf{A}_{\frac{n_T}{2}}^{j'}(\mathbf{I}-\mathbf{\Theta}_{n_T})\mathbf{A}_{\frac{n_T}{2}}^{k'}
\end{equation}
and
\begin{equation}\label{eq:EigenValuesII}
    (\mathbf{I}-\mathbf{\Theta}_{n_T})\mathbf{A}_{\frac{n_T}{2}}^{j'}(\mathbf{I}-\mathbf{\Theta}_{n_T})\mathbf{A}_{\frac{n_T}{2}}^{k'},\;(\mathbf{I}+\mathbf{\Theta}_{n_T})\mathbf{A}_{\frac{n_T}{2}}^{j'}(\mathbf{I}+\mathbf{\Theta}_{n_T})\mathbf{A}_{\frac{n_T}{2}}^{k'}\;,
\end{equation}
with ${j'}=\frac{j+1}{2}$, ${k'}=\frac{k+1}{2}$ and ${j'}\neq {k'}$. Note that the entries on the $l$th-diagonals of
the matrices $\mathbf{A}$, where $l=\pm\{1,3,\dots,n_T-1\}$ ($l=0$ represents the main diagonal, $l>0$ above the main
diagonal, and $l<0$ below the main diagonal), are equal to zero. Due to the special structure of the matrices
$\mathbf{A}$, it follows that $(\mathbf{I}\mp\mathbf{\Theta}_{n_T})\mathbf{A}_{\frac{n_T}{2}}^{j'}$ is orthogonal to
$(\mathbf{I}\pm\mathbf{\Theta}_{n_T})\mathbf{A}_{\frac{n_T}{2}}^{j'}$ and
$(\mathbf{I}\mp\mathbf{\Theta}_{n_T})\mathbf{A}_{\frac{n_T}{2}}^{j'}$ is orthogonal to
$(\mathbf{I}\mp\mathbf{\Theta}_{n_T})\mathbf{A}_{\frac{n_T}{2}}^{k'}$, with $j'\neq k'$. Thus, the eigenvalues in
(\ref{eq:EigenValuesI}) and (\ref{eq:EigenValuesII}) are zero and therefore the eigenvalues in (\ref{eq:eigenval_sum})
are independent.

 The probability density function (pdf) of the eigenvalues
$p((\mu_j^{n_T})^2)$ can be obtained from (\ref{eq:eigenval_sum}) as follows. The rank $(\mathrm{rk}(\cdot))$ of
$\mathbf{A}_2$ is $2$. Furthermore,
\begin{eqnarray}
\mathrm{rk}(\mathbf{A}_{2^n}^{j})&=& \mathrm{rk}(\mathbf{U}\mathbf{A}_{2^n}^{j}\mathbf{U}) = \mathrm{rk}\left[%
\begin{array}{cc}
  (\mathbf{I}+\mathbf{\Theta}_{n_T})\mathbf{A}_{2^{n-1}}^{j'} & \mathbf{0}  \\
  \mathbf{0} & (\mathbf{I}-\mathbf{\Theta}_{n_T})\mathbf{A}_{2^{n-1}}^{j'} \\
\end{array}%
\right] \nonumber\\
&=&\mathrm{rk}((\mathbf{I}+\mathbf{\Theta}_{n_T})\mathbf{A}_{2^{n-1}}^{j'})+\mathrm{rk}((\mathbf{I}-\mathbf{\Theta}_{n_T})\mathbf{A}_{2^{n-1}}^{j'})
=\mathrm{rk}(\mathbf{A}_{2^{n-1}}^{j'})\;, \nonumber
\end{eqnarray}
where $\mathbf{U}$ contains the eigenvectors of $\mathbf{A}_{2^n}^{j}$. Since $\mathrm{rk}(\mathbf{A}_2)=2$, the
matrices $\mathbf{A}_{n_T}^{j}$ have all rank $2$, and thus the following holds
\begin{equation}\label{eq:SVD_A_Idem}
    \mathbf{V}^H(\mathbf{A}_{n_T}^{j})\mathbf{V}=\left[%
\begin{array}{cc}
  \mathbf{I}_2 & \mathbf{0} \\
  \mathbf{0} & \mathbf{0} \\
\end{array}%
\right]\;,
\end{equation}
where $\mathbf{V}$ is a unitary matrix.
 With (\ref{eq:SVD_A_Idem}), the pdfs are given as
\begin{eqnarray}
    p((\mu_j^{n_T})^2)&=& p\left(\mathrm{tr}\left[\sum\limits_{i=1}^{n_R}\mathbf{h}_i^H\mathbf{A}_{n_T}^{j}\mathbf{h}_i\right]\right)
     = p\left(\mathrm{tr}\left[\sum\limits_{i=1}^{n_R}\mathbf{\bar{h}}_i^H\left[%
\begin{array}{cc}
  \mathbf{I}_2 & \mathbf{0} \\
  \mathbf{0 }& \mathbf{0} \\
\end{array}%
\right]\mathbf{\bar{h}}_i\right]\right)\;,\nonumber
\end{eqnarray}
which is the sum of squares of $2n_R$ independent complex normal distributed variables, i.e. a noncentral chi-square
distribution with $4n_R$ degrees of freedom.
 That concludes the proof.
\end{proof}
\section{Proof of corollary~\ref{lem:Tightness_lowerBound}}\label{sec:Proof_TighnessLowBound}
\begin{proof}
The inequality (\ref{eq:MutualInfoUpperBound}) is tight only, if the ratio of two eigenvalues, i.e.
$r=\nicefrac{\mu_i^2}{\mu_j^2}=1$, for all $i\neq j$. From this it follows that it has to be shown that the
following holds
\begin{equation}\nonumber
    \lim\limits_{n_R\rightarrow \infty}\mathrm{Pr}(r=1)=1\;.
\end{equation}
Since the eigenvalues are chi-square distributed with each $4n_R$ degrees of freedom, the ratio of the eigenvalues
is distributed as follows
\begin{equation}\nonumber
    h(r,n_r)=\frac{\Gamma(4n_R)}{\Gamma(2n_R)^2}\frac{r^{\nicefrac{(4n_R-2)}{2}}}{(1+r)^{4n_R}}\;,
\end{equation}
which is the well-known F distribution~\cite[p.365, eq.9.9.35]{Springer}. Therefore, when $n_R$ goes infinity, the
F distribution is given as
\begin{equation}\nonumber
    \lim\limits_{n_R\rightarrow \infty}(h(r,n_R))=\delta(r-1)\;,
\end{equation}
where $\delta$ is the delta distribution. It follows that the lower bound gets tight for increasing $n_R$. The lower
bound is also tight for low SNR values, which is obvious after rewriting (\ref{eq:MutualInfoQSTBCdet}) as follows
\begin{equation}\nonumber
I_Q=\frac{2}{n_T}\log_2\left(\left(1 +\frac{\rho}{n_T}\alpha_1\right)^{\frac{n_T}{2}}-\zeta \right)\;.
\end{equation}
Furthermore, $(1 +\frac{\rho}{n_T}\alpha_1)^{\frac{n_T}{2}}\gg\zeta$ for small SNR. As an example,
$\zeta=(\frac{\rho}{n_T}\alpha_2)^2$ for $n_T=4$. Therefore, as the SNR gets smaller, the lower bound gets tighter.
That concludes the proof.
\end{proof}

\end{document}